\documentclass[twocolumn]{aastex63}

\usepackage{natbib}
\usepackage{hyperref}
\def\deg{\ifmmode^\circ\else$^\circ$\fi}

\shorttitle{New picture of IC 5146 dark Streamer}
\shortauthors{L.~K. Dewangan et al.}

\begin{document}

\title{IC 5146 dark Streamer: is a first reliable candidate of edge collapse, hub-filament systems, and intertwined sub-filaments?}

\correspondingauthor{L.~K. Dewangan}
\email{Email: lokeshd@prl.res.in}

%\author[0000-0002-7367-9355]{A.~K. Maity}
%\affiliation{Astronomy \& Astrophysics Division, Physical Research Laboratory, Navrangpura, Ahmedabad 380009, India}
%\affiliation{Indian Institute of Technology Gandhinagar Palaj, Gandhinagar 382355, India}

\author[0000-0001-6725-0483]{L.~K. Dewangan}
\affiliation{Astronomy \& Astrophysics Division, Physical Research Laboratory, Navrangpura, Ahmedabad 380009, India}

\author[0000-0001-8812-8460]{N.~K.~Bhadari}
\affiliation{Astronomy \& Astrophysics Division, Physical Research Laboratory, Navrangpura, Ahmedabad 380009, India}
\affiliation{Indian Institute of Technology Gandhinagar Palaj, Gandhinagar 382355, India}

\author{A.~Men'shchikov}
\affiliation{Université Paris-Saclay, Université Paris Cité, CEA, CNRS, AIM, 91191, Gif-sur-Yvette, France}

\author[0000-0003-0014-1527]{E.~J.~Chung}
\affiliation{Department of Astronomy and Space Science, Chungnam National University, Daejeon, Republic of Korea}

\author{R.~Devaraj}
\affiliation{Dublin Institute for Advanced Studies, 31 Fitzwilliam Place, Dublin D02XF86, Ireland}

\author{C.~W.~Lee}
\affiliation{Korea Astronomy and Space Science Institute, 776 Daedeokdae-ro, Yuseong-gu, Daejeon 34055, Republic of Korea}
\affiliation{University of Science and Technology, Korea (UST), 217 Gajeong-ro, Yuseong-gu, Daejeon 34113, Republic of Korea}

\author[0000-0002-7367-9355]{A.~K. Maity}
\affiliation{Astronomy \& Astrophysics Division, Physical Research Laboratory, Navrangpura, Ahmedabad 380009, India}
\affiliation{Indian Institute of Technology Gandhinagar Palaj, Gandhinagar 382355, India}

\author{T.~Baug}
\affiliation{Satyendra Nath Bose National Centre for Basic Sciences, Block-JD, Sector-III, Salt Lake, Kolkata-700 106, India}

%\author{et al.}

%\author[0000-0003-2062-5692]{H.~Sano}
%\affiliation{National Astronomical Observatory of Japan, Mitaka, Tokyo 181-8588, Japan}

%\author[0000-0002-1411-5410]{K.~Tachihara}
%\affiliation{Department of Physics, Nagoya University, Furo-cho, Chikusa-ku, Nagoya 464-8601, Japan}

%\author{Y.~Fukui}
%\affiliation{Department of Physics, Nagoya University, Furo-cho, Chikusa-ku, Nagoya 464-8601, Japan}

%{Evidence of cloud-cloud collision in W31}
%
%\author[A.~K. Maity et al.]
%{L.~K. Dewangan$^{1}$\thanks{lokeshd@prl.res.in} et al.\\
%$^{1}$Physical Research Laboratory, Navrangpura, Ahmedabad - 380 009, India.}
%
%{A.~K. Maity$^{1,2}$\thanks{arupmaity@prl.res.in}, L.~K. Dewangan$^{1}$, et. al\\%H. Sano$^{3,4}$, Y. Fukui$^{3,4}$ and N.~K Bhadari$^{1,2}$\\ 
%, and D.~K. Ojha$^{3}$\\
%$^{1}$Physical Research Laboratory, Navrangpura, Ahmedabad - 380 009, India.\\
%$^{2}$Indian Institute of Technology Gandhinagar Palaj, Gandhinagar 382355, India.\\
%$^{3}$Institute for Advanced Research, Nagoya University, Furo-cho, Chikusa-ku, Nagoya 464-8601, Japan.\\
%$^{4}$Department of Physics, Nagoya University, Furo-cho, Chikusa-ku, Nagoya 464-8601, Japan.}
%$^{3}$Department of Astronomy and Astrophysics, Tata Institute of Fundamental Research, Homi Bhabha Road, Mumbai 400 005, India.}

%\author{L.~K. Dewangan, et al.}
%\affil{Physical Research Laboratory, Navrangpura, Ahmedabad 380009, India}

% Abstract of the paper
\begin{abstract}
The paper presents an analysis of multi-wavelength data of a nearby star-forming site IC 5146 dark Streamer (d $\sim$600 pc), which has been treated as a single and long filament, {\it fl}.
Two hub-filament systems (HFSs) are known toward the eastern and the western ends of {\it fl}. 
Earlier published results favor the simultaneous evidence of HFSs and the end-dominated collapse (EDC) in {\it fl}. 
{\it Herschel} column density map (resolution $\sim$13\rlap.{$''$}5) reveals two intertwined sub-filaments (i.e., {\it fl-A} and {\it fl-B}) toward {\it fl}, displaying a nearly double helix-like structure. This picture is also supported by the C$^{18}$O(3--2) emission. The scenario ``fray and fragment" may explain the origin of intertwined sub-filaments. 
In the direction of {\it fl}, two cloud components around 2 and 4 km s$^{-1}$ are depicted using the $^{13}$CO(1--0) and C$^{18}$O(1--0) emission, and are connected in velocity space. The HFSs are spatially found at the overlapping areas of these cloud components and can be explained by the cloud-cloud collision scenario. 
Non-thermal gas motion in {\it fl} with larger Mach number is found. 
The magnetic ﬁeld position angle measured from the ﬁlament's long axis shows a linear trend along the filament. This signature is confirmed in the other nearby EDC filaments, presenting a more quantitative confirmation of the EDC scenario.
Based on our observational outcomes, we witness multiple processes operational in IC 5146 Streamer. Overall, the Streamer can be recognized as the first reliable candidate of edge collapse, HFSs, and intertwined sub-filaments together.

%{\bf The distribution of magnetic field angle with respect to the longer axis of IC 5146 Streamer and the other three nearby EDC filaments show a linear trend. This analysis may present a more quantitative confirmation of the EDC scenario. }
%Although in site IC 5146, the contamination caused by HFSs appears to affect such configurations.
%The onset of cloud cloud collision (CCC) seems to explain the origin of HFSs in {\it fl}. 
%, with an aspect ratio greater than 5. 
%An observed oscillatory-like velocity pattern along {\it fl} supports the presence of the intertwined structures and the fragments along {\it fl}. 
%
\end{abstract}
%------------------
%
%
%%%%%%%% INTRODUCTION %%%%%%%%%%%
\section{Introduction}
\label{sec:intro}
Several studies of star-forming regions (SFRs) have revealed that dust and molecular filaments are actively involved in star formation processes. In SFRs, a system of filaments converging to a central hub, which is referred to as a hub-filament system \citep[HFS;][]{myers09}, is a commonly observed feature, where young stellar objects (YSOs) and massive OB-stars (M $\gtrsim$ 8 M$_{\odot}$) are born. Furthermore, researchers have also identified isolated filaments undergoing end-dominated collapse (EDC) or edge-collapse \citep{bastien83,pon12,clarke15}, and observational evidence of such EDC filaments is gradually increasing in the literature (e.g., NGC 6334 \citep{2013A&A...554L...2Z}, Sh 2-242 \citep[or S242;][]{dewangan19x,2020A&A...637A..67Y}, IC 5146 \citep{wang19,chung22}, Monoceros R1 \citep[or Mon R1;][]{2020ApJ...899..167B}, G341.244-00.265 \citep{Yu2019A&A...622A.155Y}, and G45.3+0.1 \citep{bhadari22}). 
Using multi-wavelength approach, a few SFRs are identified and reported, where the edge-collapse and the HFSs are simultaneously investigated (e.g., G45.3+0.1 \citep{bhadari22}; IC 5146 dark Streamer \citep{wang19,chung22}). 
In such cases, an HFS is observed toward each edge of the EDC filament. In the HFS, filaments are identified with large aspect ratio \citep[i.e., filament's length/filament's width, A$>$5;][]{andre10} compared to the hub region ($<$3).
%(i.e., A=$\frac{filament's~ major~axis}{filament's~minor~axis}>$5 compared to the hub region ($<$3).}
 It strongly suggests the onset of the range of star formation processes within a single filamentary cloud and their connection. 
In addition to these two highlighted configurations, observational evidences for the twisting/coupling of filaments (or intertwined filaments)
have also been reported in only a couple of star-forming regions (e.g., NGC 6334 filament \citep{shimajiri19}, Lynds Bright Nebulae \citep{dewangan21}). 
However, the simultaneous signature of the HFSs, EDC, and twisted/intertwined filaments in a single SFR is not yet observed.
At present, we do not know if there exists any connection/association between these three observational configurations of filaments (i.e., HFSs, EDC, and twisted nature), which is essential to understand the formation of YSOs and massive stars.
In this context, the target of this work is the IC 5146 dark Streamer, which is a nearby promising site (d $\sim$600 pc) to search for different observational configurations of filaments. 
%However, the simultaneous study of the HFSs, EDC, and twisted/intertwined filaments in a single SFR is not yet performed. 
%At present, we do not know if there exists any connection between these three observational configurations of filaments (i.e., HFSs, EDC, and twisted nature) or associated multiple scenarios operational in a single physical system, which is very essential to ultimately understand the formation of YSOs and massive stars. 

This paper focuses on the IC 5146 dark Streamer/Northern Streamer/filamentary structure, which is one of the densest molecular clouds in IC~5146 \citep{herbig08,roy11,Arzoumanian2013,chung21}.
 The Cocoon Nebula and the Northern Streamer have been reported as the two main components of IC~5146 in the constellation Cygnus. Using the near-infrared (NIR) data, an extinction map of IC~5146 was produced by \citet{lada94}. 
Embedded dust clumps, filaments, and signposts of star formation activities (i.e., YSOs and outflows) have been reported toward 
the Cocoon Nebula and the Northern Streamer \citep{herbig08,harvey08,arzoumanian11,arzoumanian19,johnstone17,zhang20}.
\citet{chung21} explored several molecular lines (i.e., $^{13}$CO, C$^{18}$O, N$_{2}$H$^{+}$, HCO$^{+}$, CS, SO, 
NH$_{2}$D, and H$^{13}$CO$^{+}$) toward IC 5146 using the Taeduk Radio Astronomy Observatory (TRAO) 
14-m telescope, and examined filaments and dense cores in IC 5146 \citep[see also][]{dobashi93,lada94}. 
The clouds associated with the Cocoon Nebula and the Streamer were traced in velocity 
ranges of [6, 9] and [1, 7] km s$^{-1}$, respectively \citep{chung21}. 
Using the {\it Gaia} measurements, \citet{wang20} reported distance estimations of 800$\pm$100 pc and 600$\pm$100 pc for the Cocoon and the Streamer, respectively. 
Based on the earlier reported works, the Cocoon Nebula and the Northern Streamer can be treated as two distinct sources \citep[e.g.,][]{chung21}. 

Dust continuum maps revealed a long filamentary morphology of the IC 5146 dark Streamer \citep{kramer03}, and 
two prominent HFSs were investigated toward the eastern and the western parts (i.e., E-HFS and W-HFS) of the Streamer \citep[e.g.,][]{arzoumanian11,arzoumanian19,roy11,johnstone17,wang17,wang19,wang20,chung21,chung22}. 
Several polarimetry studies were conducted toward the 
IC 5146 dark Streamer \citep{wang17,wang19,wang20,chung22}, 
which allowed them to explore magnetic field structures. 
These earlier works showed the presence of uniform magnetic field vectors perpendicular to the dark Streamer. 
%A curved magnetic field morphology was investigated toward both the hubs in IC 5146 \citep{wang19,chung22}. 
In general, recent {\it Planck} observations \citep{Planck2016A&A...586A.138P} showed that the low column density filaments (striations; $N_{\rm H}<10^{21.7}$ cm$^{-2}$) are parallel to the Galactic magnetic field whereas the magnetic field is perpendicular to the filaments with higher column density ($N_{\rm H}>10^{21.7}$ cm$^{-2}$). These observations are consistent with the theoretical outcomes saying that the magnetic field supports the filaments against collapsing into its longer axis and guides the gravity-driven gas contraction \citep[e.g.,][]{Nakamura2008ApJ...687..354N,Inutsuka2015A&A...580A..49I}. 
%However, it is not well understood how magnetic fields (or B-fields) affect the star formation processes. 
A curved magnetic field morphology was investigated toward both ends of IC 5146 \citep{wang19,chung22}, and was proposed as an indication of the EDC process in long filaments. However, such signature is not yet assessed in potential and nearby EDC filaments.

\citet{chung21} found the supersonic nature of the E-HFS and W-HFS in IC 5146, and suggested a collision process of turbulent converging flows to explain the observed HFSs. 
Based on the positions of the observed HFSs at the long filament, the scenarios -- edge-driven collapse and accretion flows -- were proposed in the Streamer \citep[see Figure~9 in][]{chung22}.
However, a careful and thorough investigation of the collision process is yet to be done.
Furthermore, despite its proximity, the structures in the target site IC 5146 streamer have not been fully investigated. Several other nearby sites unveil the presence of multiple sub-filaments, cores as well \citep[e.g.,][]{Hacar2013,Hacar2018}.
%possibility of sub-filaments (i.e., intertwined filaments) in the target site IC 5146 Streamer cannot be ruled out.}
%{\bf Additionally, despite of its proximity, no attempt is made to explore a twisting nature of filaments (or intertwined filaments) in the target site IC 5146 Streamer. }
%In this relation, we find that the {\it Herschel} sub-millimeter images and the published TRAO molecular line data are not yet fully explored. 
In order to probe various physical processes operational in the Streamer, we have carefully examined the dust continuum maps from the {\it Herschel}
Gould Belt survey (HGBS) project \citep[e.g.,][]{andre10,arzoumanian11,arzoumanian19}, and the velocity structures using the published TRAO $^{13}$CO and C$^{18}$O line data. 
We have also employed the {\it ``getsf-hires"} algorithm \citep{getsf_2022} to produce high-resolution column density and temperature maps (resolution $\sim$13\rlap.{$''$}5). 

Various observational data sets used in this work are presented in Section~\ref{sec:obser}. 
The derived observational findings are reported in Section~\ref{sec:data}. 
Section~\ref{sec:disc} deals with a discussion of our observed outcomes. 
Finally, Section~\ref{sec:conc} presents a summary of the major findings of this study.

\section{Data Sets and Analysis}
\label{sec:obser}
In this paper, we selected an area of $\sim$0$\degr$.746 $\times$ 0$\degr$.464 (central coordinates: {\it l} = 93$\degr$.659; {\it b} = $-$4$\degr$.419) around the IC 5146 dark Streamer. 
The positions of YSOs (i.e., Class~I, Flat-spectrum, and Class~II) were collected from \citet{harvey08}. 
The published TRAO $^{13}$CO (1--0) and C$^{18}$O (1--0) line data \citep[resolution $\sim$49$''$;][]{chung21} were utilized in this work. 
This paper also uses an integrated C$^{18}$O (3--2) map obtained from James Clerk Maxwell Telescope (JCMT; proposal id: M06BGT02; rest frequency = 329.3305453 GHz). The C$^{18}$O (3--2) map is downloaded from the JCMT Science Archive/Canadian Astronomy Data Centre (CADC), which is a pipeline product (with pixel scale $\sim$7\rlap.{$''$}3 and resolution $\sim$14$''$) and was observed in the scan (raster) mode using the Heterodyne Array Receiver Programme/Auto-Correlation Spectral Imaging System \citep[HARP/ACSIS;][]{buckle09} spectral imaging system to cover a larger area of the Streamer.

The HGBS images at 70--500 $\mu$m and the filament skeletons \citep[e.g.,][]{arzoumanian11,arzoumanian19} 
were downloaded from the HGBS archive. 
The NRAO VLA Sky Survey \citep[NVSS; resolution $\sim$45$''$; 1$\sigma$ $\sim$0.45 mJy/beam;][]{condon98} 1.4 GHz 
radio continuum map and the {\it Planck} sub-millimeter map at 353 GHz or 850 $\mu$m \citep[resolution $\sim$294$''$;][]{planck14} 
were also utilized. 

We produced the {\it Herschel} H$_{2}$ column density and dust temperature maps at different resolutions (i.e., 13\rlap.{$''$}5, 18\rlap.{$''$}2, 24\rlap.{$''$}9, and 36\rlap.{$''$}3) using the {\it getsf-hires} method described in \citet{getsf_2022}. 
The {\it getsf} utilities \citep[e.g., {\it modfits} and {\it resample};][]{getsf_2022} were used in the analysis. 
The utility {\it modfits} was used to convert the {\it Herschel} flux densities at 70 and 160 $\mu$m from units of Jy pixel$^{-1}$ to MJy sr$^{-1}$, while the utility {\it resample} was utilized to regrid all the {\it Herschel} images at 70--500 $\mu$m 
to the pixel scale of the image at 70 $\mu$m (i.e., 3$''$). Thereafter, images with different possible resolutions (i.e., 8\rlap.{$''$}4, 13\rlap.{$''$}5, 18\rlap.{$''$}2, 24\rlap.{$''$}9, and 36\rlap.{$''$}3) were generated, and the final spectral fitting was performed (excluding the image at 70 $\mu$m) to generate high-resolution column density and temperature maps. 
The {\it Herschel} 70 $\mu$m image is generally not used in the spectral fitting due to the contamination caused by warm emission originating from UV-heated dust (e.g., polycyclic aromatic hydrocarbons or transiently heated small dust grains), thus leading to overestimation in dust temperature (i.e., underestimation in column density).
However, even excluding the 70 $\mu$m continuum image in the spectral fitting, the {\it getsf-hires} can still produce the images with a resolution of 8\rlap.{$''$}4 \citep[i.e., the resolution of image at 70 $\mu$m; see][for more details]{getsf_2022}. Our derived high-resolution column density and temperature maps at the resolution of 70 $\mu$m continuum image are noisy, possibly because of insignificant features present in the image at 70 $\mu$m. Therefore, in this paper, we have utilized the column density and temperature maps at resolutions of $\sim$13\rlap.{$''$}5, 18\rlap.{$''$}2, 24\rlap.{$''$}9, and 36\rlap.{$''$}3.
\section{Results}
\label{sec:data}
In this section, we present the results derived using a careful analysis of the {\it Herschel} 
data and the molecular line data, which have enabled us to uncover new insights into physical 
processes operating in the IC 5146 dark Streamer. 
\subsection{Signatures of two intertwined sub-filaments in the IC 5146 dark Streamer}
\label{subsec:maps}
\begin{figure*}
\center
\includegraphics[width=11.2 cm]{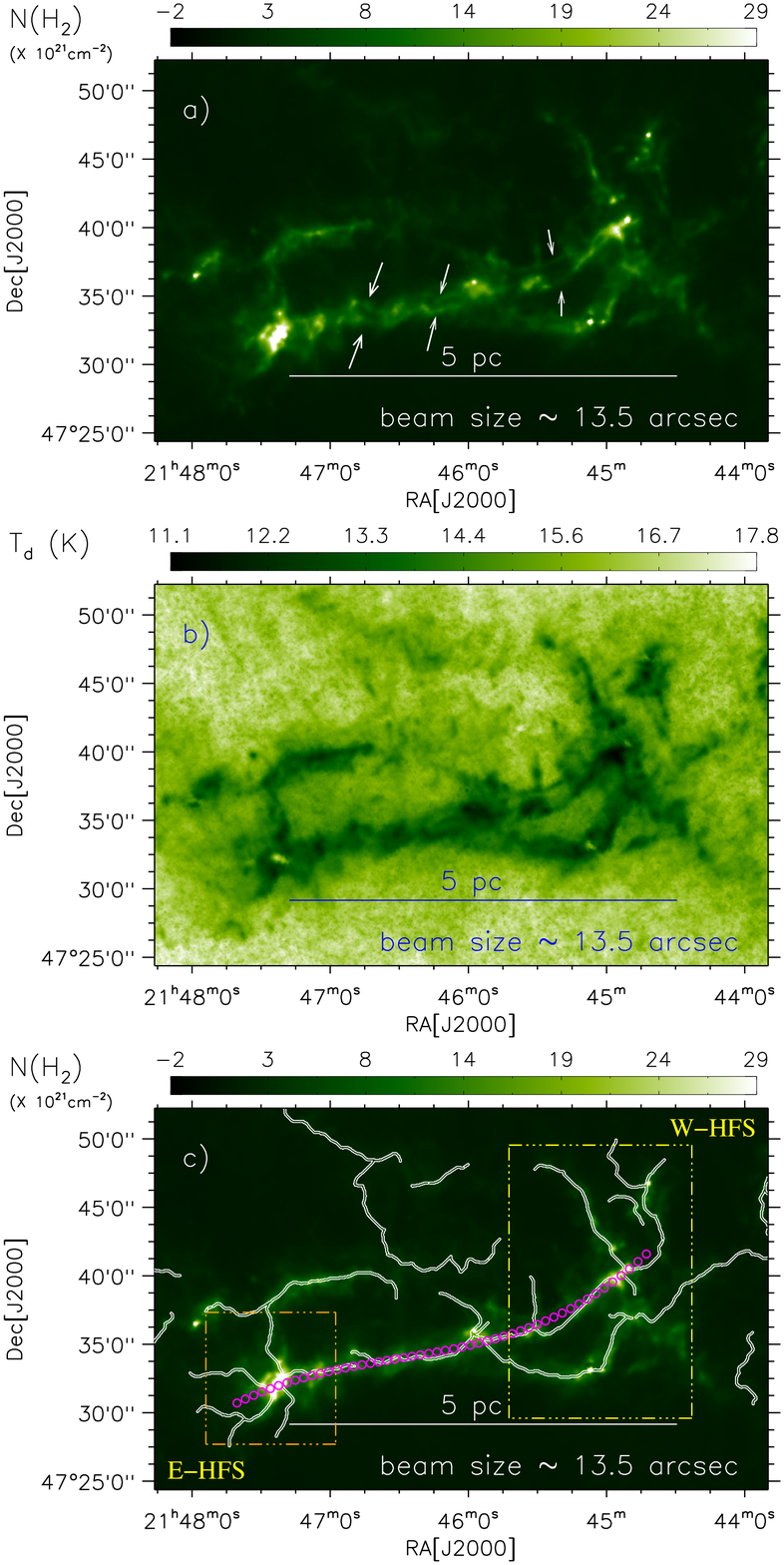}
\caption{a) The panel shows the {\it Herschel} column density map (resolution $\sim$13\rlap.{$''$}5) 
of an area containing the IC 5146 dark Streamer (size $\sim$0$\degr$.746 $\times$ 0$\degr$.464). 
Different sub-filaments are indicated by arrows. 
b) The panel presents the {\it Herschel} temperature map (resolution $\sim$13\rlap.{$''$}5).
c) Same as Figure~\ref{fig1}a, but it shows the {\it Herschel} filament skeletons \citep[see white curves from][]{arzoumanian19} and several small circular regions (in magenta; radii = 25$''$) along the filamentary structure. Two dotted-dashed boxes indicate the locations of the HFSs, which are labeled as E-HFS and W-HFS. 
A scale bar corresponding to 5 pc (at a distance of 600 pc) is shown in each panel.}
\label{fig1}
\end{figure*}

Figures~\ref{fig1}a and~\ref{fig1}b display the {\it Herschel} column density ($N(\mathrm H_2)$) and temperature ($T_\mathrm{d}$) maps (resolution $\sim$13\rlap.{$''$}5) of our selected target area around the IC 5146 dark Streamer. For the first time, we present these high resolution maps of our target site, which have been produced using the {\it getsf-hires} method as discussed in Section~\ref{sec:obser}. Visual inspection of the column density map shows several extended structures. 
Previously reported two HFSs (i.e., E-HFS and W-HFS) and filament skeletons are also shown in the {\it Herschel} column density map 
(see Figure~\ref{fig1}c), which were identified using the HGBS column density map at the resolution of $\sim$18\rlap.{$''$}2 
\citep[e.g.,][]{arzoumanian19}. In earlier works, a single long filament, designated as {\it fl}, hosting E-HFS and W-HFS at its opposite ends was mainly discussed. 
We have also marked several small circular regions (radii = 25$''$) along this long filament (see magenta circles in Figure~\ref{fig1}c), where some physical parameters (such as column density, temperature etc.) are determined (see Section~\ref{sec:coem}).
In this paper, the {\it Herschel} column density map at the resolution of $\sim$13\rlap.{$''$}5 reveals the presence of two 
intertwined sub-filaments (i.e., {\it fl-A} and {\it fl-B}; $T_\mathrm{d}$ $\sim$11--15~K), which are indicated by arrows in Figure~\ref{fig1}a. We have considered the long filament {\it fl} as the main filament, which seems to be composed of two intertwined sub-structures. 
This is a new result in the IC 5146 dark Streamer. The implication of this outcome is discussed in Section~\ref{sec:procs1}.

\begin{figure*}
\center
\includegraphics[width=17.5cm]{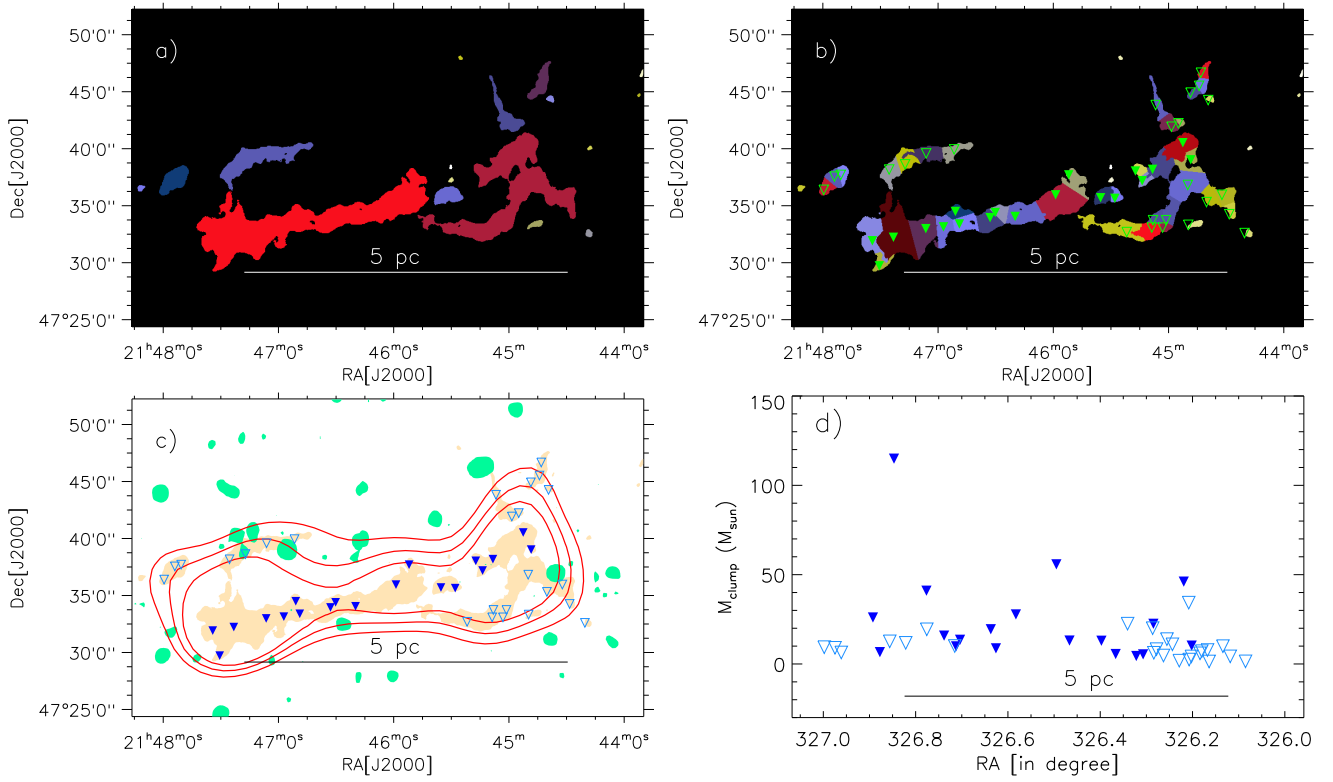}
\caption{a) Several structures are presented, and are depicted in the column density map at a contour 
level of 5.22 $\times$ 10$^{21}$ cm$^{-2}$. b) Spatial distribution of clumps identified toward the structures shown in 
Figure~\ref{fig3}a. The positions of these clumps are indicated by filled and open upside down triangles. 
The boundary of each {\it Herschel} clump is also shown. 
Clumps highlighted by filled symbols are selected toward the elongated structure (see Figure~\ref{fig1}c), 
while open symbols present clumps located away from the elongated structure. 
c) The panel shows a filled $N({{\rm{H}}}_{2})$ contour (in moccasin) at 5.22 $\times$10$^{21}$ cm$^{-2}$, 
a filled NVSS 1.4 GHz continuum contour (in spring green), and 
the emission contours at [4.92, 5.70, 6.42] MJy/sr from the {\it Planck} 353 GHz or 850 $\mu$m intensity map. 
Upside down triangles are the same as shown in Figure~\ref{fig3}b. 
d) The panel shows a variation of clump masses against their corresponding longitudes 
(see filled and open upside down triangles in Figure~\ref{fig3}b). A scale bar corresponding to 5 pc (at a distance of 600 pc) is shown in each panel.}
\label{fig3}
\end{figure*}

Figure~\ref{fig3}a shows the boundaries of different structures traced in the column density map. 
In order to identify these structures, we used the $N(\mathrm H_2)$ contour at 
5.22 $\times$ 10$^{21}$ cm$^{-2}$ and the {\it IDL} based {\it clumpfind} algorithm \citep{williams94}. 
The {\it clumpfind} algorithm divides two- and three-dimensional data into distinct emission clumps by contouring of data with a multiple of the rms noise \citep [see more information in][]{williams94}.
The identified structures primarily trace the eastern filament and the western HFS.
We have also identified several clumps toward the structures as presented in Figure~\ref{fig3}a. We used the $N(\mathrm H_2)$ contours at 
[5.22, 9.3, and 12] $\times$ 10$^{21}$ cm$^{-2}$ to trace these clumps. 
In $N(\mathrm H_2)$ map, we define the clumps as non-filamentary arbitrarily shaped structures (size $\sim$0.5 pc), while the cores
are rather circularly shaped structures (size $\leq$0.1 pc; see Section~\ref{sec:getsf}). In standard terminology, clumps have a lower density ($\sim$10$^{4}$ cm$^{-3}$) than the embedded cores \citep[$>$10$^{5}$ cm$^{-3}$;][]{Onishi2002,Saito2006}.
%() can enclose multiple cores ().
%In $N(\mathrm H_2)$ map, we define the clumps as arbitrary shaped structures (non-filamentary) can enclose multiple cores.
%While, the cores are mostly circular structures.
The locations and boundaries of the clumps are displayed in Figure~\ref{fig3}b. 
Filled upside-down triangles show the clumps distributed toward the long filament {\it fl}, 
while the clumps located away from this structure are highlighted by open upside-down triangles. 
Figure~\ref{fig3}c displays the distribution of the {\it Herschel} clumps, the ionized emission traced by the 
NVSS 1.4 GHz continuum contour, and the {\it Planck} 353 GHz or 850 $\mu$m continuum emission against 
the structures as presented in Figure~\ref{fig3}a. We do not find any noticeable radio continuum emission toward 
the long filament {\it fl}, including both the HFSs. The elongated appearance is also evident in 
the {\it Planck} 353 GHz or 850 $\mu$m continuum map, which does not show any inner structures due to its coarse beam size. 

We have also computed the mass of each {\it Herschel} clump using the equation, $M_{clump} = \mu_{H_2} m_H A_{pix} \Sigma N(H_2)$, where $\mu_{H_2}$ is the mean molecular weight per hydrogen molecule (i.e., 2.8), $A_{pix}$ is the area subtended by one pixel (i.e., 3$''$/pixel), and $\Sigma N(\mathrm H_2)$ is the total column density \citep[see also][]{dewangan17}. 
Figure~\ref{fig3}d presents the mass distribution against the position of all the clumps, allowing us to examine the mass distribution of clumps. 
Filled symbols show the clumps distributed toward the long filament {\it fl}, and massive ones appear to be present at its 
opposite ends. 

\begin{figure*}
\center
\includegraphics[width=10cm]{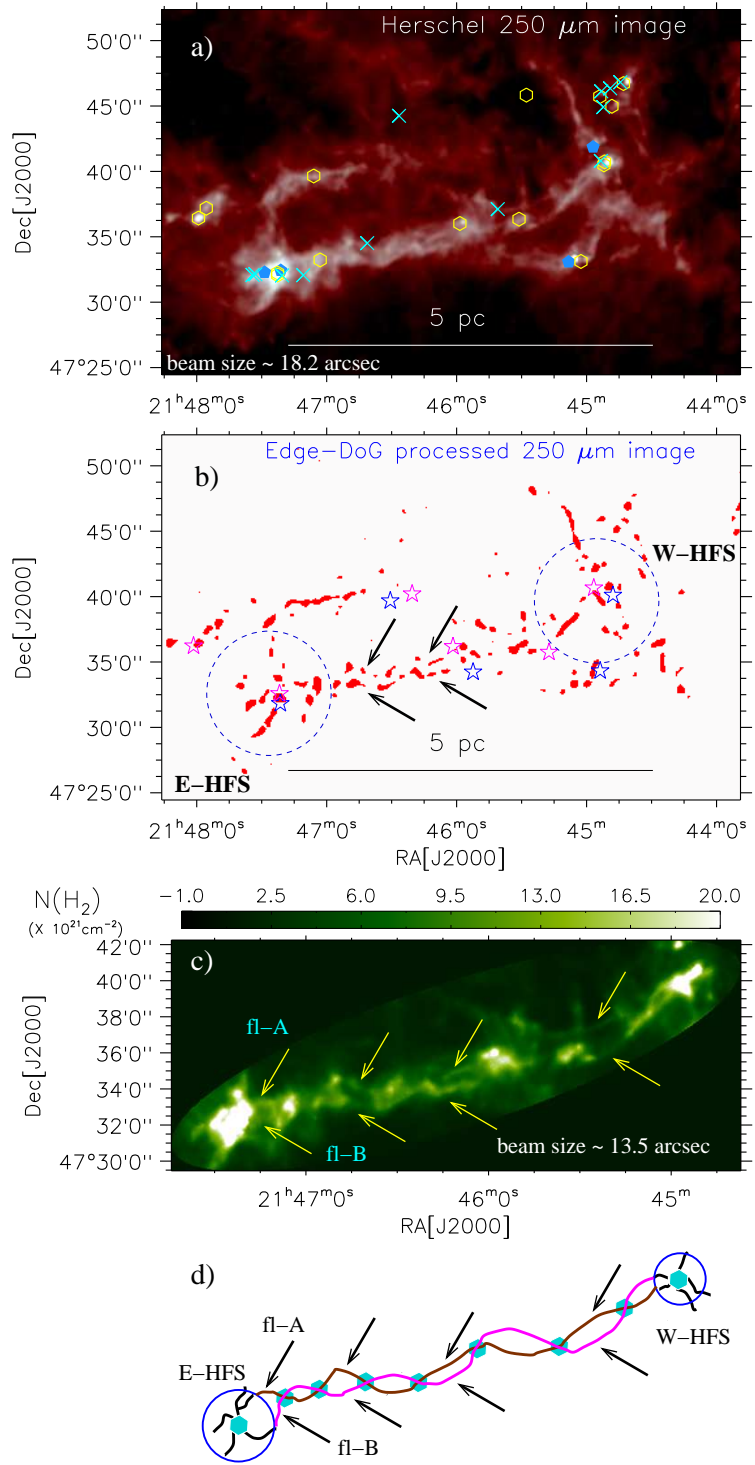}
\caption{a) Overlay of the positions of YSOs \citep{harvey08} on the {\it Herschel} image at 250 $\mu$m. 
Hexagons, filled pentagons, and multiplication signs represent Class~I YSOs, flat-spectrum sources, and Class~II YSOs, respectively. 
(b) The panel shows the {\it Herschel} image at 250 $\mu$m, which has been exposed to the ``Edge-DoG'' algorithm. 
Two sub-filaments and two HFSs (i.e., E-HFS and W-HFS) are indicated by arrows and dashed circles, respectively. 
Blue stars and magenta stars represent the blueshifted and redshifted outflow lobes \citep[from][]{zhang20}, respectively. 
c) The panel presents a zoom-in view of the {\it Herschel} column density map (resolution $\sim$13\rlap.{$''$}5) toward the IC 5146 dark Streamer. A scale bar corresponding to 5 pc (at a distance of 600 pc) is shown in panels ``a" and ``b".
d) A cartoon displaying the positions of E-HFS, W-HFS, and the possible distribution of intertwined filaments {\it fl-A} and {\it fl-B}. Filled hexagons signify the overlapping areas of {\it fl-A} and {\it fl-B}. 
} 
\label{fig4}
\end{figure*}

In order to highlight two sub-filaments, a false color map is produced using the {\it Herschel} image at 250 $\mu$m (see Figure~\ref{fig4}a). Previously reported positions of the Class~I YSOs, flat-spectrum sources, and Class~II YSOs \citep[from][]{harvey08} are shown on the {\it Herschel} image.
We also employed the ``Edge-DoG'' algorithm on the {\it Herschel} image at 250 $\mu$m, and its outcome is presented in Figure~\ref{fig4}b. 
The {\it IDL} based ``Edge-DoG'' filter enhances the extended brightness inhomogeneities (e.g., sharp edges) based on the Difference of Gaussians filters technique \citep[][]{Assirati2014}.
The positions of HFSs (i.e., E-HFS and W-HFS) and two possible sub-filaments are marked in Figure~\ref{fig4}b. 
Figure~\ref{fig4}c exhibits a zoom-in view of the {\it Herschel} column density map toward the IC 5146 dark Streamer, where the two sub-filaments {\it fl-A} and {\it fl-B} are indicated by arrows. 
Based on the visual inspection of {\it Herschel} images, we have presented a cartoon diagram displaying the possible configuration of {\it fl-A} and {\it fl-B} in Figure~\ref{fig4}d. The overlapping areas of the sub-filaments are shown by filled hexagons, where either the YSOs or dense sources are identified. The implication of this configuration is discussed in Section~\ref{sec:procs1}.

\subsection{Identification of filament skeletons and the dust continuum sources/cores on N(H$_{2}$) map at 13\rlap.{$''$}5}
\label{sec:getsf}
\begin{figure*}
\center
\includegraphics[width=14cm]{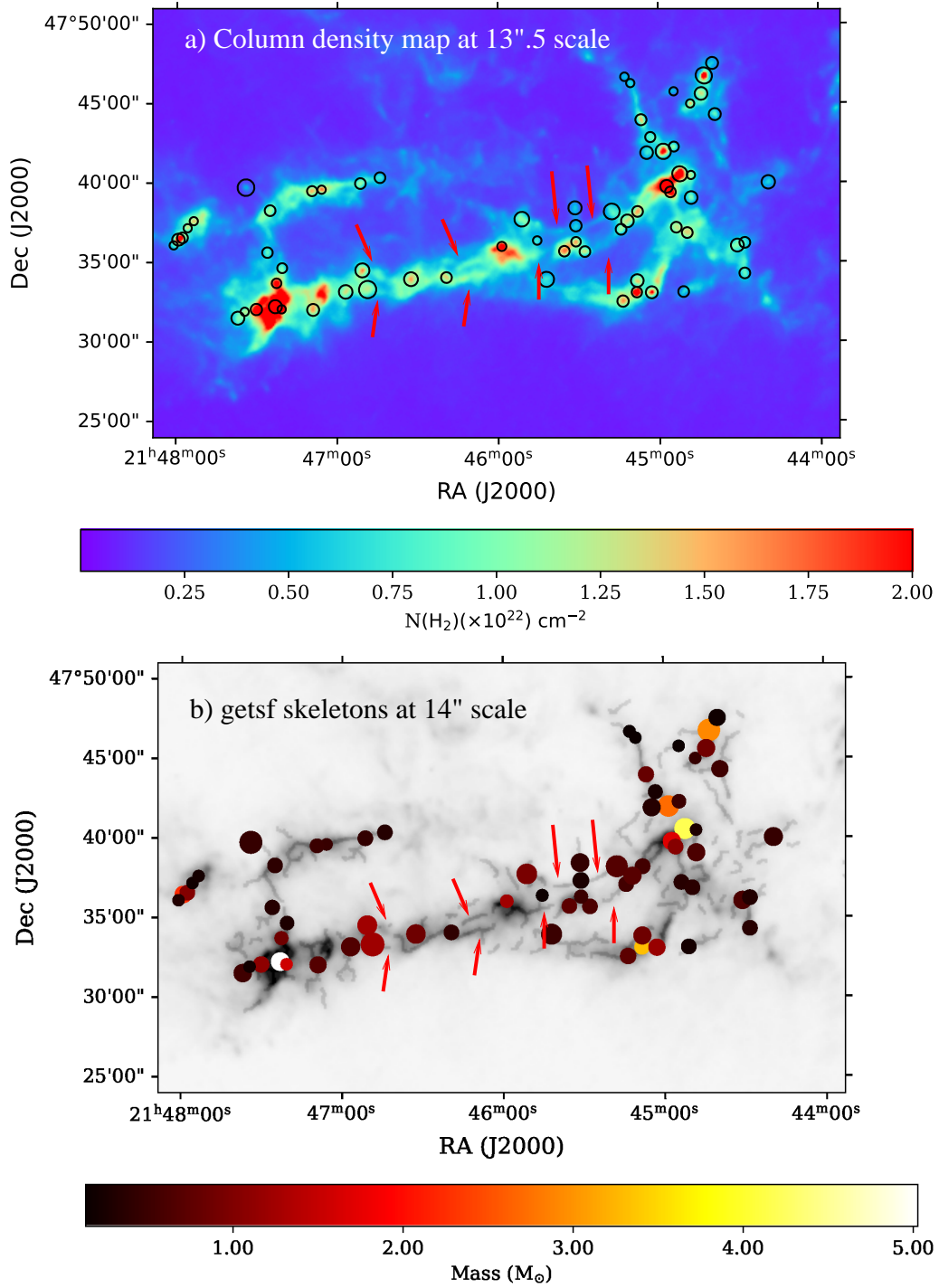}
\caption{
a) H$_{2}$ column density (N(H$_{2}$)) map at 13\rlap.{$''$}5 resolution. The {\it getsf} extracted sources are overlaid and shown by their footprint size (see text for more details).
b) Overlay of the {\it getsf} identified skeletons at 14$''$ scale on the 13\rlap.{$''$}5 N(H$_{2}$) map. 
The size of dots is proportional to the footprint area of {\it getsf} sources.
The arrows mark the sub-filaments {\it fl-A} and {\it fl-B} (see Figure~\ref{fig4}).
%c) Overlay of the {\it getsf} extracted sources on the skeletons at global scales ($\sim$14--375$''$). The overlaid sources in panels ``b" and ``c" are the same as in panel ``a", except the size indicates the relative source area (see text for more details). The color scheme represents the mass distribution of {\it getsf} sources.
} 
\label{fig4x}
\end{figure*}

To identify the dust continuum sources (or cores) and filament skeletons present in our target site IC 5146 Streamer, we used the {\it getsf} tool presented in \citet{getsf_2022}.
The {\it getsf} method is capable of extracting the sources and filaments from an astronomical image and requires only a single user input of the maximum size of the structure to extract.
We employed {\it getsf} utility on the N(H$_{2}$) map at 13\rlap.{$''$}5. The maximum source and filament size were set to be 20$''$ and 200$''$, respectively.
Figure~\ref{fig4x}a presents the overlay of {\it getsf} extracted sources on the N(H$_{2}$) map. The size of sources indicates their footprint size, which is plotted by their estimated major and minor axes.
We have only selected the sources which lie within the N(H$_{2}$) contour value of 5.22$\times$10$^{21}$ cm$^{-2}$. A total of 67 sources were found and displayed in Figure~\ref{fig4x}.
The overlay of filament skeletons on the N(H$_{2}$) map at the spatial scale of $\sim$14$''$ is shown in Figure~\ref{fig4x}b. The presence of at least two close filaments along the major axis of {\it fl} is marked by arrows. We have overlaid the {\it getsf} sources on the image. The size of dots is proportional to the footprint area of {\it getsf} sources. The color scheme of sources infers their mass distribution. 
The mass of the sources is estimated by the same method presented in Section~\ref{subsec:maps}.
This analysis suggests that the more massive cores are located at the hub locations (i.e., E-HFS and W-HFS) compared to the other areas.
%The global scale filament skeletons (scale $\sim$14--375$''$) identified by {\it getsf} are presented in Figure~\ref{fig4x}c. Global scale skeletons trace filamentary structures on all scales independent of filament width. The {\it getsf} traces filaments from the image resolution scale to the maximum scale of interest of the user. 
We cleaned the {\it getsf} identified filament skeletons by removing the spurious structures and the structures outside of our target region.

\subsection{Kinematics of molecular gas}
\label{sec:coem} 
\subsubsection{Variation of radial velocity and Mach number along the filament {\it fl}}
\label{subsec3}
\begin{figure*}
\center
\includegraphics[width=13cm]{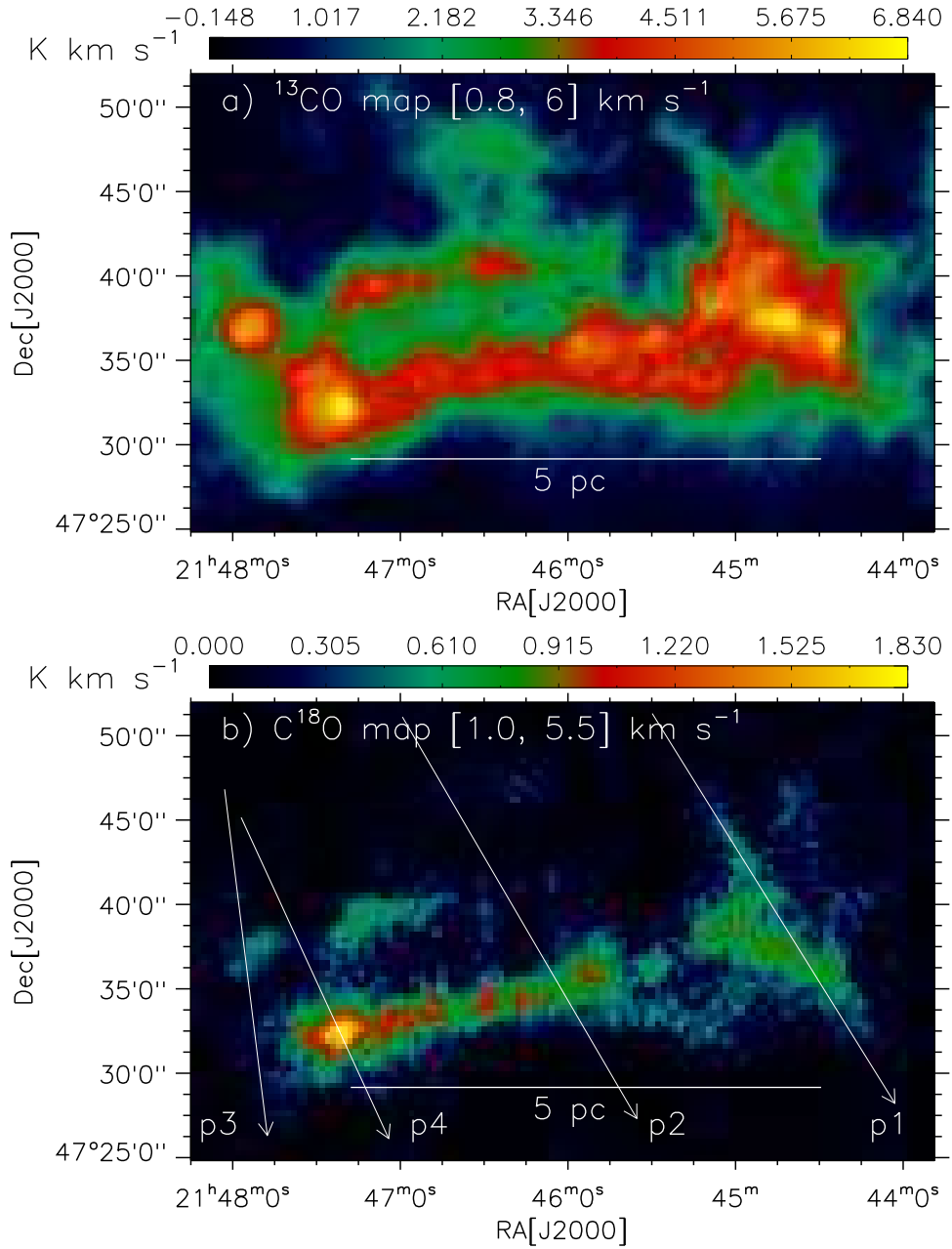}
\caption{TRAO integrated intensity (or moment-0) maps of a) $^{13}$CO and b) C$^{18}$O toward the IC 5146 dark Streamer. 
The molecular emission is integrated over a velocity interval, which is given in each panel (in km s$^{-1}$).
In panel `b', four arrows (p1--p4) are indicated, and p-v diagrams along these paths are presented in Figure~\ref{fig7}. A scale bar corresponding to 5 pc (at a distance of 600 pc) is shown in each panel.}
\label{fig5}
\end{figure*}

We have re-examined the published TRAO $^{13}$CO and C$^{18}$O line data \citep[effective beam size $\sim49''$;][]{chung21} toward the IC 5146 dark Streamer. 
Figures~\ref{fig5}a and~\ref{fig5}b display the integrated intensity (i.e., moment-0) maps of $^{13}$CO(J =1$-$0) and C$^{18}$O(J =1$-$0) 
emission over velocity ranges of [0.8, 6] and [1.0, 5.5] km s$^{-1}$, respectively. 
The distribution of both the molecular emissions seems to closely follow the long filament {\it fl}, but the two sub-filaments (i.e., {\it fl-A} and {\it fl-B}) are not spatially resolved in the molecular 
maps due to their coarse beam sizes. Both molecular emissions are more intense towards the HFSs (i.e., E-HFS and W-HFS).

\begin{figure*}
\center
\includegraphics[width=15.5 cm]{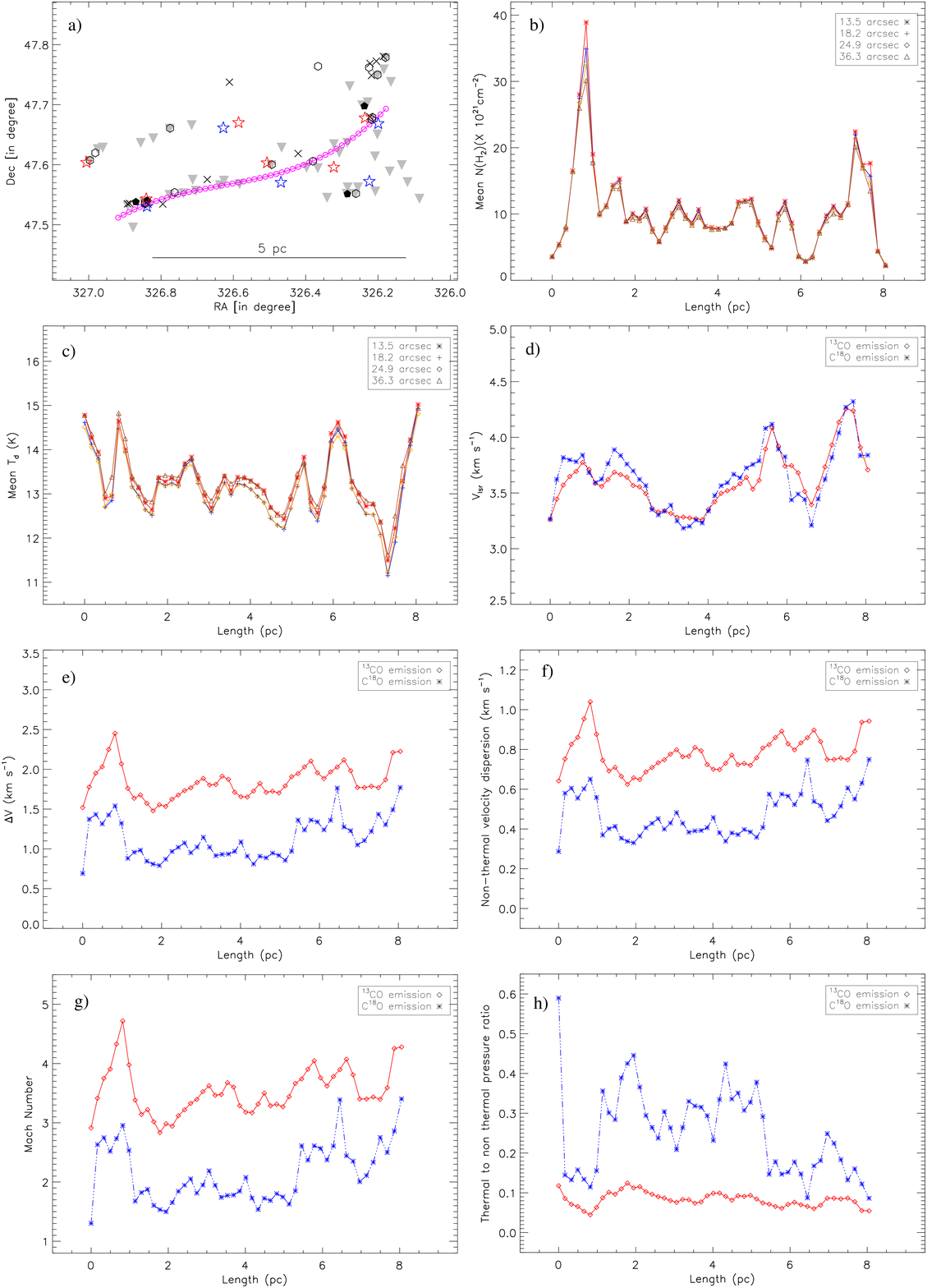}
\caption{a) The panel presents the distribution of {\it Herschel} clumps (filled upside down triangles), infrared excess sources \citep[i.e., Class~I YSOs (hexagons), flat-spectrum sources (filled pentagons), Class~II YSOs (multiplication signs); from][]{harvey08}, 
%(i.e., Class~I YSOs (hexagons), flat-spectrum sources (filled pentagons), Class~II YSOs (multiplication signs)), 
and outflow lobes \citep[stars; from][see also Figures~\ref{fig3} and~\ref{fig4}]{zhang20}. Open circles (in magenta) represent the positions of several selected regions along the filamentary structure (see also Figure~\ref{fig1}c), where average molecular spectra, average column densities, and average dust temperatures are computed. 
b--f) The panel displays a variation of the column density, dust temperature, radial velocity, FWHM line width, non-thermal velocity dispersion, Mach number, ratio of thermal to non-thermal gas pressure along the filamentary structure highlighted in Figure~\ref{fig6}a. 
In the direction of each circle marked in Figure~\ref{fig6}a, the values of dust temperature and column density are computed from the {\it Herschel} temperature and column density maps having different resolutions, while other physical parameters are extracted using the $^{13}$CO and C$^{18}$O line data.} 
\label{fig6}
\end{figure*}

In Figure~\ref{fig6}a, we show the distribution of {\it Herschel} clumps, YSOs \citep[from][]{harvey08}, and outflow lobes \citep[from][]{zhang20} toward the IC 5146 dark Streamer. We also mark open circles to highlight the positions of several circular regions along the long filament {\it fl} (see also Figure~\ref{fig1}). 
Average column density and average dust temperature are computed for each circular region using the {\it Herschel} column density and temperature maps at different resolutions (i.e., 13\rlap.{$''$}5, 18\rlap.{$''$}2, 24\rlap.{$''$}9, and 36\rlap.{$''$}3), respectively. With the help of these values, Figures~\ref{fig6}b and~\ref{fig6}c display the column density profile and the dust temperature profile along the filamentary structure, respectively. We find the column density peaks at the ends of the long filament {\it fl}, and the presence of cold dust emission ($T_\mathrm{d}$ $\sim$11--15~K) toward the filament {\it fl}. 
These outcomes are consistent in all the {\it Herschel} column density and temperature maps at different resolutions.

We also produced average spectra of $^{13}$CO and C$^{18}$O toward each circular region, enabling us to determine the mean velocity ($V_{\rm lsr}$) and the Full-Width Half Maximum (FWHM) line width ($\Delta V$) of each spectrum. 
The $V_{\rm lsr}$ and $\Delta V$ are estimated by the Gaussian fitting of each averaged spectrum.
In Figures~\ref{fig6}d and~\ref{fig6}e, we present the $^{13}$CO and C$^{18}$O velocity profiles and the $^{13}$CO and C$^{18}$O line width profiles along the filamentary structure, respectively. The variation of $V_{\rm lsr}$ and $\Delta V$ along the filament is seen. 
In other words, there is a hint of velocity oscillation along the filament. 
The knowledge of the line width value is also used to compute the non-thermal velocity dispersion, Mach number, and ratio of thermal to non-thermal pressure toward each circular region. Mach number is defined as the ratio of non-thermal velocity dispersion ($\sigma_{\rm NT}$) to sound speed ($c_{s}$). 
The sound speed $c_{s}$ (= $(k T_{kin}/\mu m_{H})^{1/2}$) can be estimated using the value of gas kinetic temperature (T$_{kin}$) and mean molecular weight ($\mu$=2.37; approximately 70\% H and 28\% He by mass). Here, we used the average dust temperature instead of T$_{kin}$, which has been estimated to be $\sim$14.2~K from the {\it Herschel} temperature map. Based on the recommendations of \citet{lada03}, we have computed the ratio of thermal to non-thermal (or turbulent) pressure (i.e., $R_{p} = {c_s^2}/{\sigma^2_{NT}}$). The non-thermal velocity dispersion can be determined using the following equation:
\begin{equation}
\sigma_{\rm NT} = \sqrt{\frac{\Delta V^2}{8\ln 2}-\sigma_{\rm T}^{2}} ,
\label{sigmanonthermal}
\end{equation}
where $\Delta V$ is defined earlier, $\sigma_{\rm T}$ (= $(k T_{kin}/30 m_H)^{1/2}$ for C$^{18}$O and = $(k T_{kin}/29 m_H)^{1/2}$ for $^{13}$CO) is the thermal broadening at T$_{kin}$ (or $T_\mathrm{d}$ $\sim$14.2~K). 
Figures~\ref{fig6}f,~\ref{fig6}g, and~\ref{fig6}h display the variation of the non-thermal velocity dispersion, Mach number, and ratio of thermal to non-thermal gas pressure for $^{13}$CO and C$^{18}$O along the filamentary structure, respectively. 
Using the $^{13}$CO and C$^{18}$O emissions, the Mach number is found to be larger than 1, and the low ratios of thermal to non-thermal 
pressure ($<$ 1) is evident for all the selected circular regions. It implies the presence of supersonic and non-thermal motion in 
the filamentary structure. Furthermore, higher values of $N(\mathrm H_2)$, Mach number, and lower values of $R_{p}$ are found toward the central hub of the E-HFS compared to the other HFS (i.e., W-HFS). 

\subsubsection{Signatures of two velocity components toward the filament {\it fl}}
\label{subsec3}
\begin{figure*}
\center
\includegraphics[width=\textwidth]{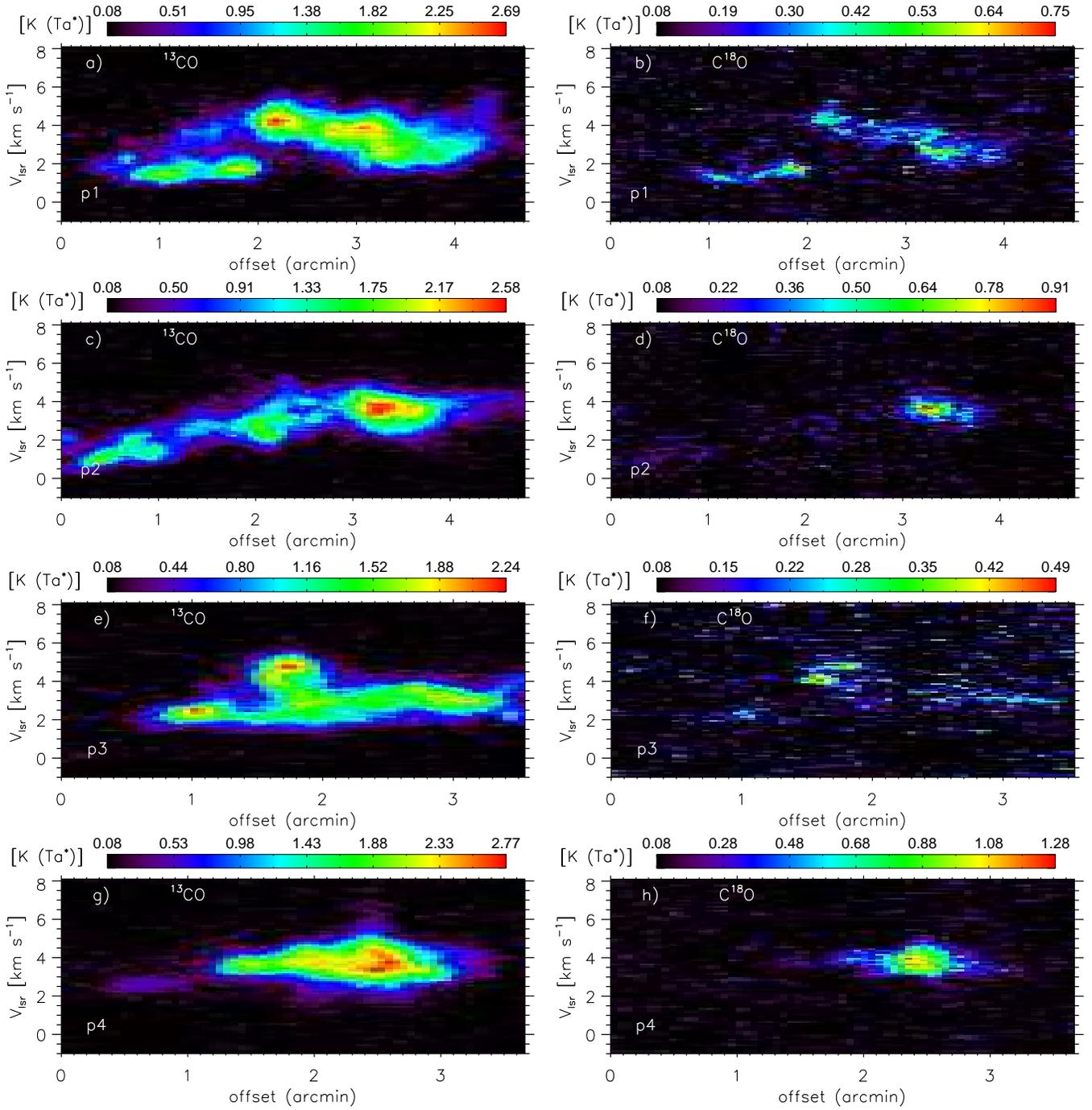}
\caption{Left column (a, c, e, g): p-v diagrams of $^{13}$CO along four arrows ``p1--p4". 
Right column (b, d, f, h): p-v diagrams of C$^{18}$O along four arrows ``p1--p4". 
These arrows ``p1--p4" are indicated in Figure~\ref{fig5}b.}
\label{fig7}
\end{figure*}

\begin{figure*}
\center
\includegraphics[width=16cm]{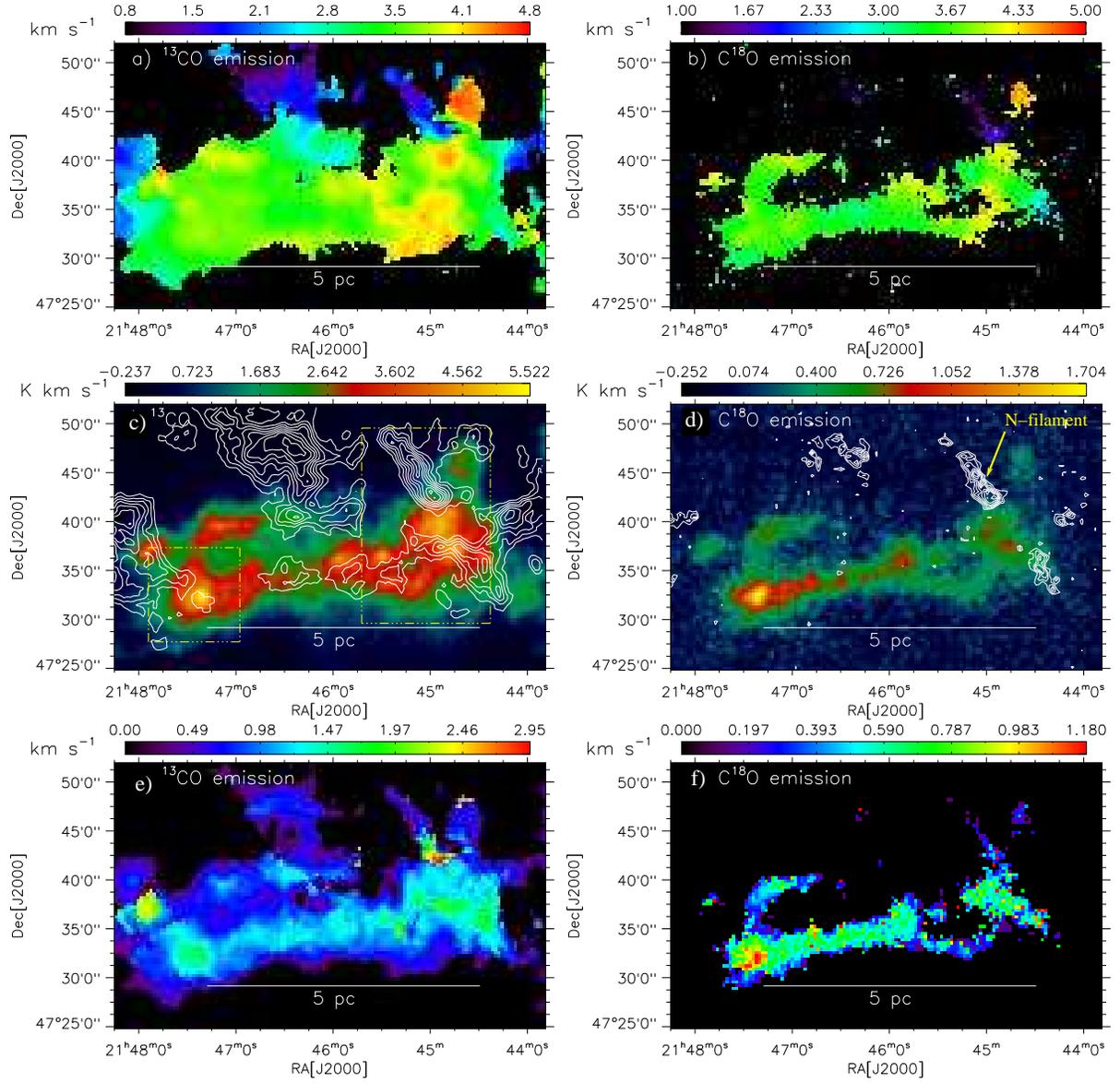}
\caption{a) $^{13}$CO moment-1 map of our selected target area (Figure~\ref{fig1}a). 
b) C$^{18}$O moment-1 map. c) Overlay of the $^{13}$CO integrated intensity emission contours at [0.8, 2.5] km s$^{-1}$ 
on the $^{13}$CO integrated intensity map at [3, 6] km s$^{-1}$. 
The contour levels are 2.56 K (in T$_{a}$*) km s$^{-1}$ $\times$ (0.2, 0.3, 0.4, 0.5, 0.6, 0.7, 0.8, 0.9).
d) Overlay of the C$^{18}$O integrated intensity emission contours at [1, 2.5] km s$^{-1}$ 
on the C$^{18}$O integrated intensity map at [3, 6] km s$^{-1}$. 
The contour levels are 0.45 K (in T$_{a}$*) km s$^{-1}$ $\times$ (0.3, 0.4, 0.5, 0.6, 0.7, 0.8, 0.9). 
e) $^{13}$CO moment-2 map. f) C$^{18}$O moment-2 map. 
A scale bar corresponding to 5 pc (at a distance of 600 pc) is shown in each panel.}
\label{fig9}
\end{figure*}

We have examined the position-velocity (p-v) diagrams of the $^{13}$CO and C$^{18}$O emissions along several lines passing through the shorter axis of {\it fl}.
In this paper, we have only presented the p-v diagrams along four arrows p1--p4 (see Figure~\ref{fig5}b).
%Position-velocity diagrams of the $^{13}$CO and C$^{18}$O emissions are produced along four arrows p1--p4 (see Figure~\ref{fig5}b). 
One can notice that the paths/arrows p1 and p4 pass through the HFS-W and HFS-E, respectively.
The p-v diagrams of $^{13}$CO along p1, p2, p3, and p4 are presented 
in Figures~\ref{fig7}a,~\ref{fig7}c,~\ref{fig7}e, and~\ref{fig7}g, respectively. 
Figures~\ref{fig7}b,~\ref{fig7}d,~\ref{fig7}f, and~\ref{fig7}h present p-v diagrams of 
C$^{18}$O along p1, p2, p3, and p4, respectively. 
In the direction of both the HFSs, p-v diagrams hint at the presence of two velocity components around 2 and 4 km s$^{-1}$, which are connected in the velocity space. 
We have also studied the velocity channel maps of the $^{13}$CO emission (not shown in this paper), supporting the existence of two velocity components (i.e., [1, 2] and [3, 4] km s$^{-1}$).
%Figure~\ref{fig8} shows the integrated velocity channel maps of the $^{13}$CO emission (at velocity intervals of 1 km s$^{-1}$), supporting the existence of two velocity components (see panels at [1, 2] and [3, 4] km s$^{-1}$). 
In this context, we show the intensity weighted mean velocity (i.e., moment-1) maps of $^{13}$CO and C$^{18}$O emission in Figures~\ref{fig9}a and~\ref{fig9}b, respectively. 

Based on a careful examination of the molecular line data, 
we have produced integrated intensity maps for two different velocity ranges (see Figures~\ref{fig9}c and~\ref{fig9}d). 
The $^{13}$CO emission maps at [0.8, 2.5] and [3, 6] km s$^{-1}$, and the C$^{18}$O emission maps at [1, 2.5] and [3, 6] km s$^{-1}$ have been generated (see Figures~\ref{fig9}c and~\ref{fig9}d). 
Figures~\ref{fig9}e and~\ref{fig9}f display the intensity weighted velocity dispersion (i.e., moment-2) maps of $^{13}$CO and C$^{18}$O emission, respectively. 
In Figure~\ref{fig9}c, we find the overlapping zones of the two cloud components, where both the HFSs and the central part of the long filament {\it fl} are spatially depicted (see also the C$^{18}$O emission in Figure~\ref{fig9}d). In these locations, higher velocity dispersions are also found, inferring the presence of either multiple velocity components (see Figures~\ref{fig9}c and \ref{fig9}e) or the presence of non-thermal gas motions (see Figures~\ref{fig9}d and \ref{fig9}f). 
Together, the molecular line data support the physical connection of the two cloud components (around 2 and 4 km s$^{-1}$) in both physical and velocity space.

\subsubsection{Existence of two intertwined molecular filaments toward the filament {\it fl}?}
\label{subsec3xx}
\begin{figure*}
\center
\includegraphics[width=\textwidth]{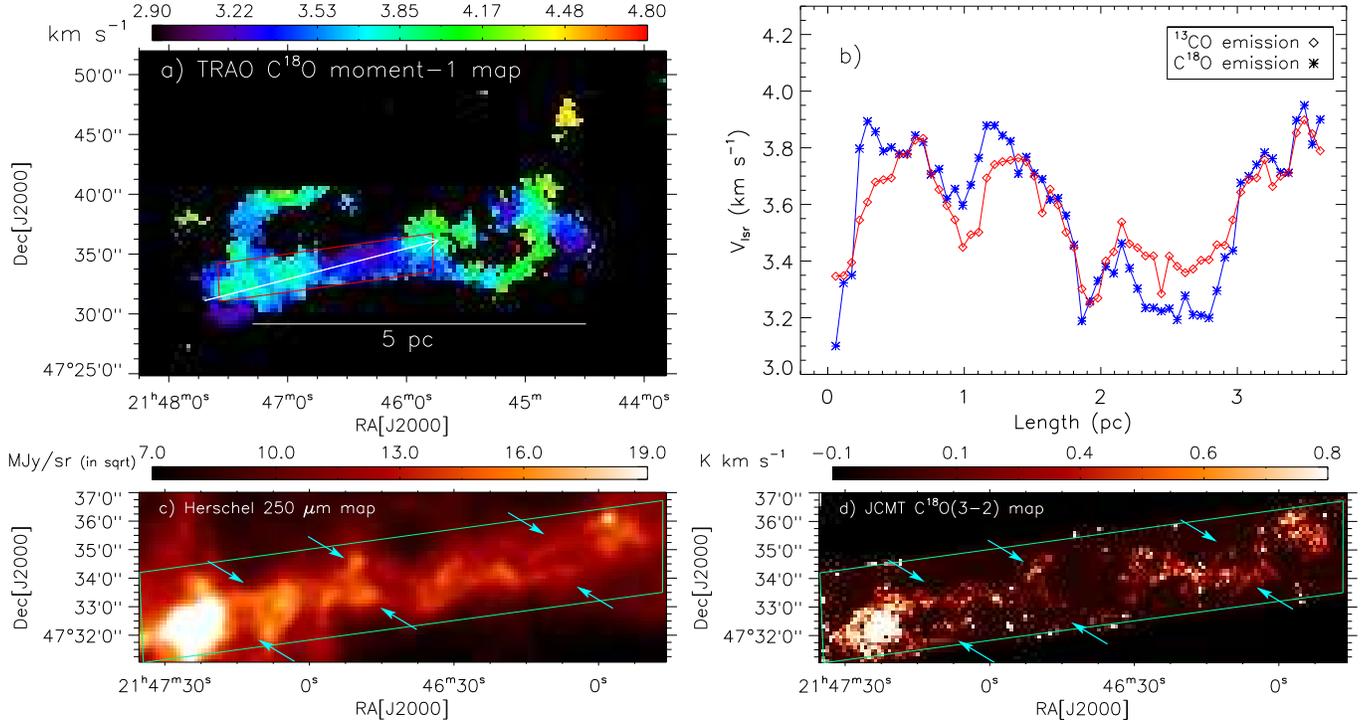}
\caption{a) The panel shows the TRAO C$^{18}$O moment-1 map of our selected target area (Figure~\ref{fig1}a), which is generated for the molecular cloud 
at [2.9, 6] km s$^{-1}$ (see Figure~\ref{fig9}f). A scale bar corresponding to 5 pc (at a distance of 600 pc) is presented. b) The panel displays a variation of radial velocities along an arrow marked in Figure~\ref{fig10x}a, which are extracted using the $^{13}$CO and C$^{18}$O moment-1 maps. 
c) The panel shows areas toward the eastern and central parts of the IC 5146 dark Streamer using the {\it Herschel} 250 $\mu$m 
image. d) The panel presents an integrated JCMT C$^{18}$O (3--2) map. In panels ``a", ``c", and ``d", a solid box highlights the area covered in the JCMT C$^{18}$O (3--2) map.}% as presented in Figure~\ref{fig10x}d.}
\label{fig10x}
\end{figure*}

The cloud component at [3, 6] km s$^{-1}$ (or around 4 km s$^{-1}$) traced using both the molecular lines has an elongated appearance similar to the long filament {\it fl}, while the second component around 2 km s$^{-1}$ does not show any elongated morphology like {\it fl} (see Figure~\ref{fig9}d). 
The previously reported N-filament \citep[see Figure~\ref{fig9}d in this paper and Figure~2 in][]{chung22} toward the western end of {\it fl} is associated with the cloud around 2 km s$^{-1}$. 
%The cloud component at [3, 6] km s$^{-1}$ (or around 4 km s$^{-1}$) traced using both the molecular lines has an elongated appearance as seen in the {\it Herschel} maps. 
Concerning the cloud component around 4 km s$^{-1}$, Figure~\ref{fig10x}a shows the moment-1 map of the C$^{18}$O emission, revealing the presence of velocity variations along the filamentary feature. To further infer velocity variations, using the moment-1 maps of $^{13}$CO and C$^{18}$O, we extracted the velocity profiles along an arrow highlighted in Figure~\ref{fig10x}a (see Figure~\ref{fig10x}b). 
A velocity variation/oscillation is traced in both molecular emissions. In the direction of the eastern and central parts of the IC 5146 dark Streamer, the {\it Herschel} image at 250 $\mu$m and the integrated JCMT C$^{18}$O (3--2) map are presented in Figures~\ref{fig10x}c and~\ref{fig10x}d, respectively. Due to the limited sensitivity, the existing JCMT C$^{18}$O (3--2) map could only hint at the existence of two intertwined molecular filaments (see arrows in Figure~\ref{fig10x}d) as investigated in the {\it Herschel} maps (i.e., {\it fl-A} and {\it fl-B}). 
For comparison purposes, we also mark arrows over the {\it Herschel} image at 250 $\mu$m. 
To further study these molecular filaments, one will need new high-resolution and high-sensitivity molecular line data for a larger area around the IC 5146 dark Streamer. 
We found that the high-resolution C$^{18}$O maps (resolution $\sim$15--20$''$) at different transitions (i.e., J=1--0, J=2--1, and J=3--2) of IC 5146 dark Streamer presented in \citet{Graham_2008} reveal the similar morphology as seen in the {\it Herschel} maps. Thus, the continuum and the molecular maps hint at the existence of two intertwined sub-filaments.

\section{Discussion}
\label{sec:disc}
Based on several previously published works, the dark Streamer of IC 5146 has been considered as a single and long filament {\it fl} having an aspect ratio larger than 5 \citep{arzoumanian11,arzoumanian19,johnstone17,wang17,wang19,wang20,zhang20,chung21,chung22}, which is also indicated in Figure~\ref{fig1}c. 
One prominent HFS has been traced toward the east-end (E-HFS) and the west-end (W-HFS) of the dark Streamer (see Figure~\ref{fig1}c). 
\citet{chung22} presented a cartoon showing the observed configuration of the Streamer (see Figure~9 in their paper).
In our high resolution {\it Herschel} column density map, both the HFSs are associated with the regions of high column densities (see Figure~\ref{fig6}b). 
%However, the E-HFS has been evident with a higher column density than the W-HFS, which was also reported in earlier works \citep[e.g.,][]{arzoumanian11,arzoumanian19,chung21,chung22}. 
Noticeable YSOs and outflow lobes \citep[e.g.,][]{harvey08,zhang20} have been reported toward both the ends of the dark Streamer (i.e., E-HFS and W-HFS; see Figure~\ref{fig6}a), 
supporting the ongoing star formation activities \citep[e.g.,][]{chung21,chung22}. 
The absence of radio continuum emission and high column density areas toward the E-HFS and W-HFS hints at the potential sites of future massive star formation (see Figure~\ref{fig3}). Earlier published results together were interpreted in favour of the edge-driven collapse and fragmentation scenario in the dark Streamer \citep[see][and references therein]{wang19,chung22}.
%Considering all these results, the edge-driven collapse and fragmentation scenario was proposed in the dark Streamer \citep[see][and references therein]{wang19,chung22}.
Additionally, in support of the proposed edge collapse scenario, the curved B-field morphology in core-scale HFSs toward both ends was detected \citep[see][and references therein]{wang19,chung22}. 
Apart from the IC 5146 Streamer, we have verified this signature in the other nearby EDC filaments as well.

%In the introduction section, we have also highlighted candidate filaments of edge collapse. Apart from the IC 5146 dark Streamer (d $\sim$600 pc), we find other three EDC filaments (i.e., Mon R1 (d $\sim$760 pc), S242 (d $\sim$2.1 kpc), and NGC 6334 (d $\sim$1.3 kpc)), which are nearby EDC systems (d $\lesssim$ 2 kpc). 
%
\subsection{Magnetic Field orientations in nearby candidate EDC filaments}
\label{sec:magdisc}
In the introduction section, we have highlighted the candidate filaments experiencing edge collapse. Apart from the IC 5146 dark Streamer (d $\sim$600 pc), we find three other EDC filaments (i.e., Mon R1 (d $\sim$760 pc), S242 (d $\sim$2.1 kpc), and NGC 6334 (d $\sim$1.3 kpc)), which are nearby EDC filamentary systems (d $\lesssim$ 2 kpc).
The existing observations of dust polarised emission from the {\it Planck} telescope have been employed to 
study the large-scale magnetic field of the four EDC filaments.
We used the {\it Planck} 353 GHz Stokes I, Q, and U maps to estimate the linear polarization angles (PAs) of polarized dust emission caused by the anisotropic dust grains in our target sites.
The stokes I, Q, and U maps were converted from the cosmic microwave background (CMB) temperature (K$_{\rm cmb}$) scale to MJy sr$^{-1}$ using the unit conversion factor of 246.54 \citep[e.g.,][]{Planck2016A&A...594A...8P}.
We also smoothed the stokes maps by {\it astropy} based Gaussian 2D-kernel (input parameter x\_stddev=2) to increase the signal-to-noise ratio.
We estimated the PAs in Galactic coordinates using the conventional relation of $\theta_{\rm GAL}=0.5\times~ {\rm arctan2(-U, Q)}$, where $-$U is used to follow the IAU convention \citep[see more details in][]{Planck2015A&A...576A.104P} and a two-argument function arctan2 is used to avoid the $\pi$-ambiguity in the estimation of PAs. The magnetic field orientations were then computed by adding 90$\degr$ in the electric field polarization PAs \citep[e.g.,][]{Planck2016A&A...586A.136P,Planck2016A&A...594A..19P}. We term this angle as B$_{\rm Gal}$ throughout the paper, which is measured from galactic north to east along the counter-clockwise direction.

\begin{figure*}
\includegraphics[width=\textwidth]{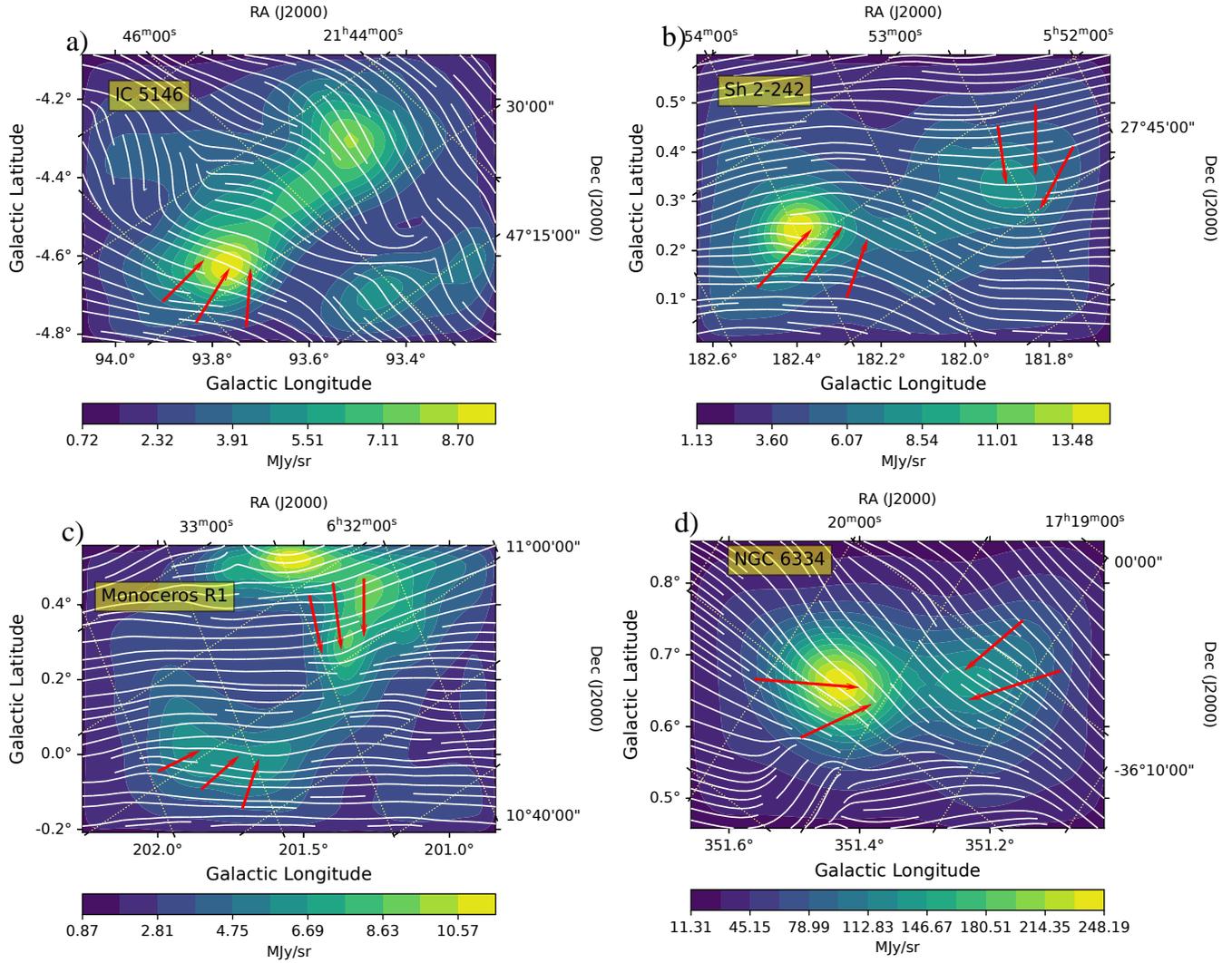}
\caption{{\it Planck} 353 GHz I-stoke images of EDC filaments IC 5146, Sh 2-242, Monoceros R1, and NGC 6334 overlaid by streamlines displaying POS magnetic field. The streamlines are displayed using {\it streamplot} in {\it matplotlib} for a density parameter of 1.0. 
The red arrows indicate the B-field curvature at the filament edges.
The Stokes Q and U maps were used to calculate the POS magnetic field position angles (see text for more details).} 
\label{ffig1}
\end{figure*}

The distribution of the plane of sky (POS) magnetic field in the direction of our selected EDC filaments IC 5146 dark Streamer, S242, Mon R1, and NGC 6334 is displayed by streamlines in Figure~\ref{ffig1}. We used {\it streamplot} in {\it matplotlib} to display the magnetic field orientations toward our selected targets. 
The magnetic field direction is nearly perpendicular to all the filaments, consistent with the observations of other targets \citep[e.g.,][]{Palmeirim2013,Planck2016A&A...586A.138P,Cox2016A&A...590A.110C}. 
The $^{13}$CO emission at the velocity range of [0.8, 2.5] km s$^{-1}$ (see Figure~\ref{fig9}c) appears to be parallel to the B-field orientation shown in Figure~\ref{ffig1}a.
It is reported that faint filaments (striations) are well aligned to the magnetic fields, and the main filament can gain its mass from these striations \citep[e.g.,][]{Palmeirim2013,zhang20}. Therefore, the $^{13}$CO emissions parallel to the B-field show that IC 5146 dark Streamer is not isolated but interacts with the natal clouds.
However, interestingly the B-field direction is curved at the edges of our target filaments. As discussed by \citet{wang19} and \citet{chung22}, the B-field directions in the EDC filaments make a shape of ``)'' and ``('' at the filament edges (hereafter, the bending effect). This is because of the material pile up at the edges, which itself moves toward the center of the filament \citep[e.g.,][]{clarke15}. This longitudinal motion of gaseous material from the ends to the center of the filament can have sufficient ram pressure that pinches the magnetic field lines and forms the U-shaped magnetic field morphology \citep[or bending effect; see more details in][]{Gomez2018MNRAS.480.2939G,wang19,Wang2020ApJ}.
This effect is most prominently seen in the S242 and Mon R1 filament than in the other sources (e.g., IC 5146 dark Streamer and NGC 6334; see arrows in Figure~\ref{ffig1}).
We suspect that the bending effect at one of the edges of the IC 5146 dark Streamer is not prominent because of the contamination caused by the presence of the HFS. However, still, this effect is significant toward the south-eastern clump of the IC 5146 dark Streamer \citep[see also Figure 7 of][]{chung22}.
Similarly, the NGC 6334 also shows the signature of HFSs at its edges \citep[see Figure 2 of][]{Tige2017A&A...602A..77T} and does not show the strong bending effect in Figure~\ref{ffig1}.
However, the high-resolution polarization map (beam $\sim$14$''$) of NGC 6334 by \citet[][]{Arzoumanian2021} indeed shows the bending effect at the dense eastern region (Figure~\ref{ffig1}) revealing the longitudinal gas motion along the filament. The direction of the magnetic field at the other end of the NGC 6334 is randomly oriented in the high-resolution map of \citet[][]{Arzoumanian2021}. However, our {\it Planck} magnetic field map shows the signature of curvature at this end, too (see Figure~\ref{ffig1}).
It is discussed in the literature that the magnetic field distortions can be caused by the outflow-driven shocks, feedback from expanding ionized fronts, and gravity-driven gas flows \citep[e.g.,][]{Arzoumanian2021,Eswar2021}. 
Also, recent observations signify that the curved magnetic field can be originated by the effect of gravity and the collision of clouds \citep[e.g.,][]{Wang2020ApJ,Wang2022ApJ}.
Therefore, it is quite possible that the intense star formation activity and the presence of HFSs at the filament edges can distort the initial bending effect in EDC filaments.
Although our target filaments are promising EDC candidates, their non-linearity can diminish the bending effect at low-resolution {\it Planck} magnetic field maps. Thus, we confirm that the other linear EDC filaments (if HFSs are not present) should show the bending effect.

\begin{figure*}
\includegraphics[width=\textwidth]{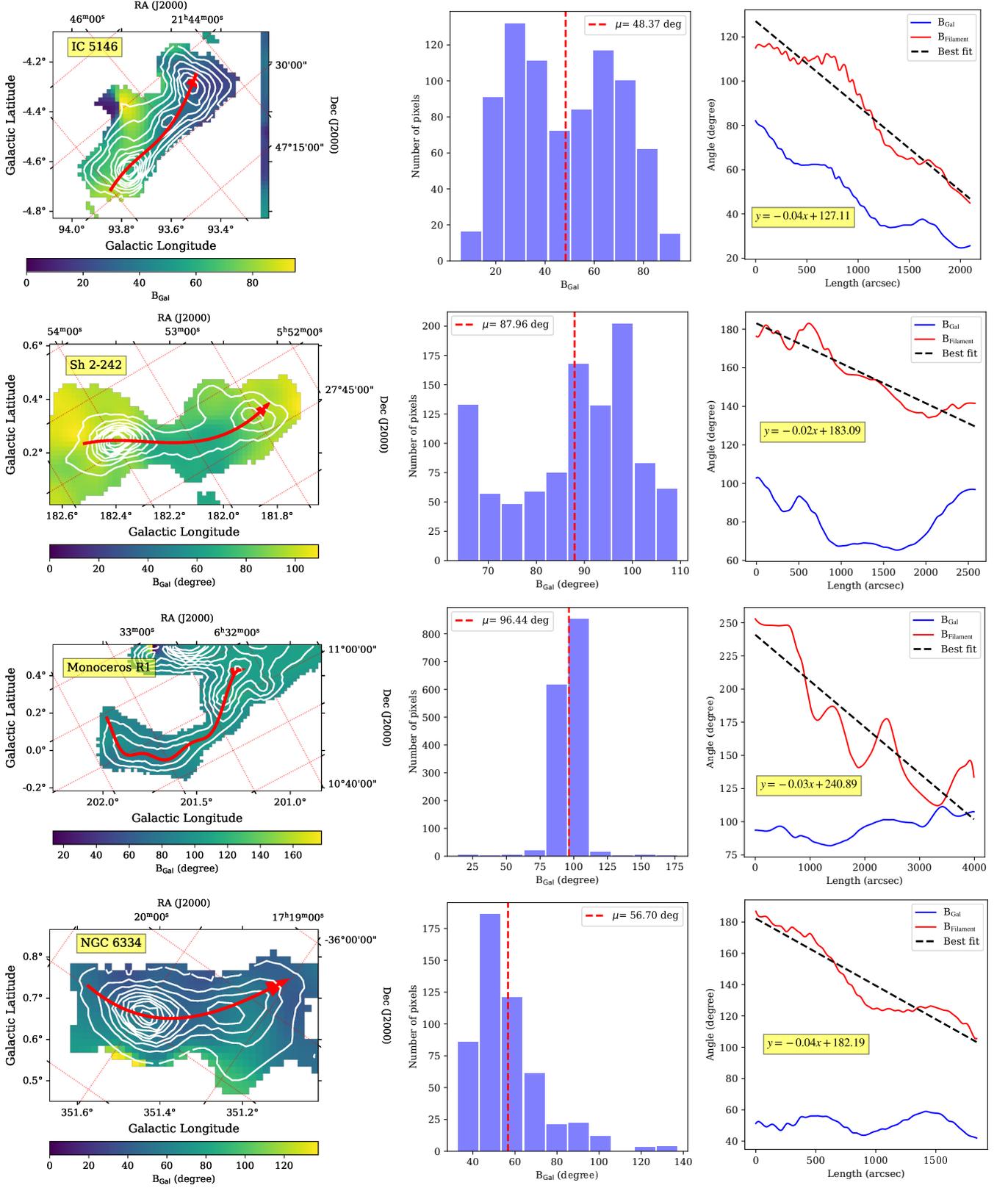}
\caption{
Distribution of the POS magnetic field angle with respect to the Galactic north to the anticlockwise direction (B$_{\rm Gal}$) toward selected EDC filaments. The first column displays the spatial distribution of B$_{\rm Gal}$ of EDC filaments IC 5146, Sh 2-242, Monoceros R1, and NGC 6334. The overlaid contours refer to the {\it Planck} 353 GHz intensity (see Figure~\ref{ffig1}). The second column presents the histograms of B$_{\rm Gal}$ corresponding to the targets shown in the first column. A vertical red dashed line indicates the mean B$_{\rm Gal}$ of histogram distribution.
The third column displays the variation of B$_{\rm Gal}$ and B$_{\rm Filament}$ along the long axis of filaments (see red curves in the first column).
B$_{\rm Filament}$ is the magnetic ﬁeld position angle measured from the ﬁlament's major axis to the anticlockwise direction (see text for more details). A best fitted line for B$_{\rm Filament}$ distribution is shown and labelled.
} 
\label{ffig2}
\end{figure*}

Figure~\ref{ffig2} displays the spatial distribution of magnetic field position angle (B$_{\rm Gal}$) toward selected EDC filaments. The derived B$_{\rm Gal}$ maps are masked out for regions of low intensity and used to obtain the histograms of B$_{\rm Gal}$ (see column~2 of Figure~\ref{ffig2}). The mean B$_{\rm Gal}$ is found to be 48.37, 87.96, 96.44, and 56.70 degrees for the IC 5146 dark Streamer, S242, Mon R1, and NGC 6334, respectively.
To derive the B$_{\rm Gal}$ distribution in the IC 5146 dark Streamer, we have considered only the region containing elongated filament. Hence we have masked out the southern-west region seen in Figure~\ref{ffig1}.
%The distribution of B$_{\rm Gal}$ along the long axis of filaments is presented by blue curve in column~3 of Figure~\ref{ffig2}. We derived the averaged B$_{\rm Gal}$ toward several circular regions of 1$'$ radius along the defined path (see red curves) and plotted them as a function of curve length. The curve length indicates the length of the filaments. A global linear gradient in B$_{\rm Gal}$ is evident for the IC 5146 dark Streamer. For the S242 and NGC 6334 filaments, B$_{\rm Gal}$ is higher at the edges compared to the central region.
Interestingly, these two filaments are nearly perpendicular to the Galactic plane. Mon R1, however, is a highly curved filament having its ends nearly perpendicular and parallel to the Galactic plane. A positive gradient in B$_{\rm Gal}$ can be seen from where the Mon R1 filament gets perpendicular to the Galactic plane.
The distribution of B$_{\rm Gal}$ along the long axis of selected filaments is presented by blue curves in Figure~\ref{ffig2}.
To quantify the distribution of the magnetic field toward the target filaments, we estimated the magnetic field position angle with respect to the filament's major axis from the direction of its head (B$_{\rm Filament}$). The distribution of B$_{\rm Filament}$ vs filament's length (B$_{\rm Filament}$--$L$) is shown by red curves in Figure~\ref{ffig2}. 
Interestingly, for all the filaments, the global trend follows a negative slope in a range of [$-$0.04, $-$0.02] degree arcsec$^{-1}$.
We estimated the slope by fitting a straight line (see black dotted line) on the B$_{\rm Filament}$--$L$ plot.
The negative slope in B$_{\rm Filament}$--$L$ distribution agrees well with the idea of magnetic field bending effect by \citet{wang19}. Ideally, a straight filament having magnetic field lines perpendicular toward its central regions and curved magnetic fields (i.e., shapes of ``)" and ``(") toward the respective edges \citep[see Figure~13 of][]{wang19}, should show the global increasing or decreasing trend in B$_{\rm Filament}$--$L$ plot.
The negative or positive trend will depend upon the choice of reference direction (i.e., filament's head or tail) from which the magnetic field angle is measured.
In our target filaments, a global linear trend in B$_{\rm Filament}$--$L$ plot hints at the EDC signature.
However, despite the linear trend, we witness local oscillations as well. This is possible because of the non-linearity of the filament (or sky-projection effect) and the contamination caused by other ongoing processes.

\subsection{Physical processes operating in the IC 5146 dark Streamer}
\label{sec:procs1}
Based on the examination of the TRAO molecular line data, a collision of turbulent converging flows was suggested to 
explain the existence of HFSs, which also includes the role of mass flow along the filaments to the dense cores in IC 5146 \citep{chung21}. 
In order to further explore the collision process in the Streamer, we revisited the TRAO $^{13}$CO and C$^{18}$O line data (see Section~\ref{sec:coem}). 
Our analysis of the molecular line data shows the existence of two cloud components around 2, and 4 km s$^{-1}$ in the direction of the IC 5146 dark Streamer, and a velocity connection of these components is also found. % The cloud around 4 km s$^{-1}$ has an elongated appearance similar to the long filament {\it fl}, while the second component around 2 km s$^{-1}$ does not show any elongated morphology like {\it fl}. 
The areas of the E-HFS and W-HFS, where the higher level of clumpiness is observed, are seen toward the common zones of the cloud components (see Section~\ref{subsec3}).
Earlier works on the cloud-cloud collision (CCC) recommend a spatial and velocity connection of two cloud components as a reliable tracer of CCC \citep[see][and references therein for more details]{fukui21,maity22}. 
In general, colliding gas flows are thought to be operated in 
the low-density medium, while the high-density phase of colliding gas flows could refer to CCC \citep[e.g.,][]{beuther20,dewangan2022new}. 

Central hubs of HFSs, hosting massive stars and clusters of YSOs, are thought to gain the inflow material from very large scales of 1-10 pc, which can be funneled along molecular filaments \citep[see][for more details]{Tige+2017,Motte+2018}. Theoretical works favor the existence of HFSs by the colliding clouds or large-scale colliding flows \citep{balfour15,inoue18}, and this scenario is also supported in the recent review article on CCC \citep{fukui21}. Based on our findings, it is likely that the IC 5146 dark Streamer may be influenced by the collision process, and both the HFSs appear to be originated in the shock-compressed interface layer by the colliding clouds/flows \citep[see also][]{chung21}. 
The massive sources are exclusively found to reside at the hub locations (see Figure~\ref{fig4x}).

Now, we have examined why HFSs are formed at the edges of the filament {\it fl}? 
It was already explained by the onset of the edge-driven collapse in the filament {\it fl}, which has a higher aspect ratio \citep[see][and references therein]{wang19,chung22}. In the EDC process, density enhancement is expected at each end 
of the filament due to high gas acceleration \citep{bastien83,pon12,clarke15,hoemann22}. 

The present work reveals the existence of two coupled (or intertwined) sub-filaments (i.e., {\it fl-A} and {\it fl-B}) toward the main filament {\it fl}, showing almost a double helix-like pattern (see Figures~\ref{fig4}d, \ref{fig4x}, and Section~\ref{subsec:maps}). 
The TRAO $^{13}$CO and C$^{18}$O line data do not reveal this configuration due to a coarse beam size. However, the JCMT C$^{18}$O (3--2) map seems to support the detected structures in the {\it Herschel} maps. Hence, such intertwined configuration is seen in both the dust and molecular emissions. 
A cartoon diagram displaying the physical configuration of sub-filaments {\it fl-A} and {\it fl-B} (Figure~\ref{fig4}d) is motivated by their simultaneous detection in the dust continuum and the molecular emission maps.
%Figure~\ref{fig4}d presents a cartoon displaying the physical configuration of {\it fl-A} and {\it fl-B}. This drawing is motivated by the similar morphology in dust continuum and molecular emission maps. 
However, the existing molecular line data do not allow us to trace the velocity information along each sub-filament. Previously, \citet{dewangan21} found the velocity oscillation along two intertwined filaments in site LBN 140.07+01.64. The physical configuration of sub-filaments in IC 5146 Streamer appears very similar to the sub-filaments identified in LBN 140.07+01.64 (see Figures~3 and 7 therein).
%seems to match with sub-filaments identified in LBN 140.07+01.64 (see Figures~3 and 7 therein). 
Therefore, due to such limited sites in literature, the presence of intertwined filaments or their double helix-like pattern and their role in star formation demand more observational as well as theoretical insights.
The sub-filaments {\it fl-A} and {\it fl-B} appear to spatially overlap with each other along the major axis, forming multiple common areas where the {\it Herschel} clumps and YSOs are seen. 
This is another new outcome of this work. In this relation, we examined the existing ``fray and fragment" scenario of the formation of intertwined sub-structures \citep{tafalla15,clarke17}, 
which firstly predicts the formation of the main filament by a collision of two supersonic turbulent gas flows, and then the scenario favors the origin of the intertwined system of velocity-coherent sub-structures in the main filament due to residual turbulent motions and self-gravity \citep[see also][]{smith14,shimajiri19}. This scenario was proposed in the NGC 6334 filament \citep{shimajiri19} and the Taurus filament \citep{tafalla15}.

Additionally, we have also investigated a noticeable velocity oscillation along the filament {\it fl} (see Section~\ref{subsec3}). Previously, in the case of a filament G350.5-N associated with the cloud G350.54+0.69, \citet{liu19} reported 
a large-scale periodic velocity oscillation. To explain this aspect, on the basis of gravitational-instability-induced core formation 
models, they proposed a proposal with the combination of longitudinal gravitational instability and a large-scale physical oscillation along the filament. Considering two sub-filaments and the distribution of the {\it Herschel} clumps toward the long filament {\it fl}, the observed velocity oscillations/variations seem to support the presence of two coupled or intertwined sub-filaments and fragment/clump formation along the filament, where the non-thermal (or turbulent) pressure seems to be dominated (see Section~\ref{subsec3}).

Taken together, our findings reveal the onset of multiple physical processes in the IC 5146 dark Streamer, which includes the edge collapse, CCC, accretion flows, and ``fray and fragment" scenario. 
\section{Summary and Conclusions}
\label{sec:conc}
In order to probe ongoing physical processes in a nearby star-forming site IC 5146 dark Streamer (d $\sim$600 pc), we have conducted a study using the multi-wavelength data. The dark Streamer resembles a single and long filament, {\it fl}, having an aspect ratio $>$ 5. 
The eastern and the western ends of {\it fl} harbor one HFS. Simultaneous detections of the HFSs and the edge-collapse have been reported in the filament {\it fl}. High resolution {\it Herschel} column density map (resolution $\sim$13\rlap.{$''$}5) is produced in this work and shows higher column densities toward both the HFSs. 
%But, the W-HFS is associated with relatively lower column density than the E-HFS.
The {\it Herschel} column density map also displays two intertwined sub-filaments (i.e., {\it fl-A} and {\it fl-B}) toward the main filament {\it fl}. 
Such configuration displays almost a double helix-like pattern, which is also seen in the integrated intensity map of the JCMT C$^{18}$O(3--2) emission. 

Using the TRAO $^{13}$CO(1--0) and C$^{18}$O(1--0) line data cubes, we have found the existence of two cloud components (around 2 and 4 km s$^{-1}$) toward the main filament. The cloud component around 4 km s$^{-1}$ has an elongated appearance like the main filament {\it fl}. Both the HFSs are spatially seen at the common regions of the cloud components. 
The origin of HFSs in {\it fl} may be explained by the CCC process. A careful analysis of the $^{13}$CO and C$^{18}$O emission shows the presence of non-thermal motion in {\it fl} with a larger Mach number. The central hub of the E-HFS shows higher values of $N(\mathrm H_2)$, Mach number, and lower values of $R_{p}$ compared to the W-HFS. 
The study of velocity profiles along the filament {\it fl} shows an oscillatory-like velocity pattern, favoring the presence of the intertwined structures and the fragments along {\it fl}. 
The origin of the intertwined sub-structures in {\it fl} seems to be explained by the scenario ``fray and fragment". 
The study of the {\it Planck} polarimetric maps of potential and nearby EDC filaments (d $\lesssim$ 2 kpc) supports the detection of a curved magnetic field morphology as a signature for the edge collapse as discussed in \citet{wang19}. 
In IC 5146 dark Streamer, we suspect that the bending effect of the magnetic field toward the western hub may be diminished by the presence of evolved HFS (i.e., W-HFS). 
The magnetic ﬁeld position angle measured from the ﬁlament's major axis shows a linear trend along the filament. This signature is confirmed in the other nearby EDC filaments. Thus, this analysis may present a more quantitative observational proxy of the EDC scenario.
Taking into account all our derived results, the IC 5146 dark Streamer can be considered the first reliable candidate of edge collapse, HFSs, and intertwined sub-filaments. 
\section*{Acknowledgments}
We thank the anonymous reviewer for the constructive comments, which have improved the scientific content of the manuscript.
The research work at Physical Research Laboratory is funded by the Department of Space, Government of India. 
C.W.L. is supported by the Basic Science Research Program through the National Research Foundation of Korea (NRF) funded by the Ministry of Education, Science and Technology (NRF-2019R1A2C1010851), and by the Korea Astronomy and Space Science Institute grant funded by the Korea government (MSIT; Project No. 2022-1-840-05).
D. R. has been supported by the European Research Council advanced grant H2020-ER-2016-ADG-743029 under the European Union’s Horizon 2020 Research and Innovation program.
NKB thanks Jia-Wei Wang for providing the useful comments on the manuscript.
This research has made use of data from the Herschel Gould Belt survey (HGBS) project (http://gouldbelt-herschel.cea.fr). The HGBS is a Herschel Key Programme jointly carried out by SPIRE Specialist Astronomy Group 3 (SAG 3), scientists of several institutes in the PACS Consortium (CEA Saclay, INAF-IFSI Rome and INAF-Arcetri, KU Leuven, MPIA Heidelberg), and scientists of the Herschel Science Center (HSC).
This research made use of {\it Astropy}\footnote{http://www.astropy.org}, a community-developed core Python package for Astronomy \citep{astropy2013,astropy2018,astropy:2022} and {\it matplotlib} \citep{Hunter:2007} Python package.

%\par
%\textbf{Facilities: NANTEN2}

%%%%%%%%%%%%%%%%%%%% REFERENCES %%%%%%%%%%%%%%%%%%

\bibliographystyle{aasjournal}
\bibliography{reference}{}
%\nocite{*}

%%%%%%%%%%%%%%%%%%%%%%%%%%%%%%%%%%%%%%%%%%%%%%%%%%

%
%\begin{figure*}
%\center
%\includegraphics[width=\textwidth]{fg7.eps}
%\caption{Integrated velocity channel maps of $^{13}$CO emission (at velocity intervals of 1 km s$^{-1}$). A scale bar corresponding to 5 pc (at a distance of 600 pc) is shown in each panel.} 
%\label{fig8}
%\end{figure*}
%
%%%%%%%%%%%%%%%%%%%%%%%%%%%%%%%%%%%%%%%%%%%%%%%%%%%%%%%%%%%%%%%%%%%%%%%%%%%%%%%%%%%%%%%%%%%%

\end{document}